%% file: main.tex
\newif\ifarXiv
\arXivtrue 

\documentclass[aoas]{imsart}

\RequirePackage{amsthm,amsmath,amsfonts,amssymb}
\RequirePackage[authoryear]{natbib}
\RequirePackage[colorlinks,citecolor=blue,urlcolor=blue]{hyperref}
\RequirePackage{graphicx}

\usepackage{orcidlink}
\startlocaldefs

\endlocaldefs

\begin{document}

\begin{frontmatter}
\title{A scalable Bayesian framework for Galaxy emission line detection and redshift estimation}
\runtitle{Bayesian emission line detection}

\begin{aug}
\author[A]{\fnms{Alexander}~\snm{Kuhn\,\orcidlink{0009-0000-1707-7133}}\ead[label=e1]{kuhn0218@umn.edu}},
\author[B]{\fnms{Bonnabelle}~\snm{Zabelle\,\orcidlink{0000-0002-7830-363X}}\ead[label=e2]{kusch037@umn.edu}},
\author[A]{\fnms{Sara}~\snm{Algeri}\ead[label=e3]{salgeri@umn.edu}},
\author[A]{\fnms{Galin}~\snm{L. Jones\,\orcidlink{0000-0002-6869-6855}}\ead[label=e4]{galin@umn.edu}},
\and
\author[B]{\fnms{Claudia}~\snm{Scarlata\,\orcidlink{0000-0002-9136-8876}}\ead[label=e5]{mscarlat@umn.edu}}
\address[A]{School of Statistics, University of Minnesota\printead[presep={,\ }]{e1,e3,e4}}

\address[B]{School of Physics and Astronomy, University of Minnesota\printead[presep={,\ }]{e2,e5}}
\end{aug}

\begin{abstract} 
Estimating galaxy redshifts is crucial for constraining key physical quantities like those in the equation of state of dark energy. Modern telescopes such as the James Webb Space Telescope, the Euclid Space Telescope, and the NASA Nancy Grace Roman Space Telescope are producing massive amounts of spectroscopic data that enable precise redshift estimation. However, a galaxy's redshift can be estimated only when emission lines are present in the observed spectrum, which is unknown {\em a priori}. A novel Bayesian approach to estimating redshift and simultaneously testing for the presence of emission lines is developed. Although modern spectroscopic surveys involve millions of spectra and give rise to highly multimodal posterior distributions, the proposed framework remains computationally efficient, admitting a parallelizable implementation suitable for large-scale inference.
\end{abstract}


\end{frontmatter}

\section{Introduction}
Despite significant progress in understanding the large-scale structure and evolution of the universe, major uncertainties remain about the nature of dark energy and dark matter. These components continue to elude direct detection and challenge our understanding of fundamental physics.

Several ongoing and planned experiments aim to address these uncertainties. The Euclid space telescope (a European Space Agency mission with contributions from NASA) and the future NASA Nancy Grace Roman Space Telescope are two such missions, designed to map the geometry of the dark universe and investigate the cause of its accelerating expansion. Both will construct three-dimensional maps of the universe by measuring galaxy \emph{redshifts}, which serve as proxies for distance. As the universe expands, light from more distant galaxies is stretched to longer wavelengths, and redshift quantifies this stretching. Because the precise relationship between distance and redshift depends on the underlying cosmological model, these measurements can place strong constraints on the nature of dark energy \citep{laureijsEuclidDefinitionStudy2011, euclidCollaborationEuclidOverview2025}.

High-precision galaxy redshifts are obtained from \emph{spectroscopic data}, which measure light intensity across wavelengths to produce \emph{spectra}. Traditional slit-based spectroscopy works by isolating a single galaxy's light with a narrow slit and dispersing only that selected beam. To observe many galaxies simultaneously, Euclid and Roman instead use \emph{slitless spectroscopy}, which disperses light from all galaxies in a telescope's field of view at once, producing spectra for numerous galaxies in a single exposure. This high-throughput approach enables large surveys, but introduces additional challenges for determining galaxy redshifts. Spectra can overlap, and the observed light at a given wavelength may include contributions unrelated to any particular galaxy. Slitless spectroscopy also suffers from a lower signal-to-noise ratio and reduced spectral resolution compared to slit-based methods \citep{baronchelliIdentificationSingleSpectral2020}. 

Within these spectra, the most informative features for redshift estimation are bright peaks produced by elements such as hydrogen and oxygen, called \emph{emission lines}. These lines have known \emph{rest-frame} wavelengths -- the wavelengths they would occur at if the galaxy were stationary relative to the observer. In observed spectra, emission lines appear at higher wavelengths than their rest-frame values due to the galaxy's redshift, but otherwise retain their characteristic shapes (see Section~\ref{sec:lines} for details). Importantly, not every spectrum contains emission lines, and their presence and observed locations are not known \emph{a priori}. While even a single line can provide information about a galaxy's redshift, detecting multiple lines greatly improves precision, especially in the presence of spectral overlap and in cases with lower signal-to-noise ratios. A central aim of this work is to provide principled uncertainty quantification for redshift estimation that accounts for this variability in emission-line presence and strength.

Roman and Euclid provide slitless spectroscopic capabilities over large areas of the sky (on the order of half a square degree per pointing). The James Webb Space Telescope (JWST) offers a complementary approach, performing slitless spectroscopy with its Near Infrared Imager and Slitless Spectrograph (NIRISS) and Near Infrared Camera (NIRCam) instruments. JWST covers smaller fields of view (a few thousandths of a square degree) at lower spectral resolution, but reaches substantially greater depths than Euclid and Roman. Despite these differences in survey area and depth, the fundamental challenges in measuring spectroscopic redshifts remain similar. 

In this manuscript, we present an analysis of data from JWST as a testbed; however, the methodology is designed with the large survey volumes expected from Euclid and Roman in mind. Accordingly, our approach prioritizes computational efficiency to enable the analysis of millions of spectra.

\subsection{Main challenges and existing work}
We develop a Bayesian approach to simultaneously detect multiple emission lines and estimate redshift that explicitly accommodates the most common systematic errors induced by slitless spectroscopic data. To the best of the authors' knowledge, no other existing approach has done so.

Most redshift estimation pipelines currently rely on \emph{template fitting}, typically through cross-correlation maximization \citep{tonrySurveyGalaxyRedshifts1979, zamoraRevisingCrossCorrelation2024} or $\chi^2$-minimization \citep{boltonSpectralClassificationRedshift2012}. These methods match the observed spectrum to astrophysical templates constructed from simulations, high-quality observations, or both. While simple and fast, template-fitting methods do not provide a principled framework for emission line detection or for quantifying redshift uncertainty.

In an attempt to address the limitations of traditional template fitting, \cite{jamalAutomatedReliabilityAssessment2018} construct a Bayesian model based on a set of known templates and utilize Bayes factors to test for the presence of emission lines. This provides posterior redshift estimates and a principled method for line detection when the available set of templates adequately represents the observed spectra. However, a relevant template may not exist for every spectrum, introducing unaccounted model uncertainty and increasing the risk of false line detections. Our work addresses these issues by using a template-free model.

Other template-free approaches exist. However, they are limited to searching for a single emission line rather than multiple lines simultaneously. These include machine learning-based classification methods \citep{baronchelliIdentificationSingleSpectral2020, baronchelliIdentificationSingleSpectral2021} and hierarchical Bayesian models \citep{parkSearchingNarrowEmission2008, harrisonRedshiftsGalaxiesRadio2017}. These Bayesian approaches typically rely on Markov Chain Monte Carlo (MCMC) for posterior inference, which can be especially challenging in this setting. Even when searching for a single emission line, the posterior can be multimodal, as illustrated by  \cite{parkSearchingNarrowEmission2008} in X-ray spectroscopy. When considering multiple lines simultaneously, the likelihood is non-smooth and more severely multimodal due to the changing set of observable emission lines for different redshift values: a challenge not addressed in previous work.

\begin{figure}
    \centering
    \includegraphics[width=\linewidth]{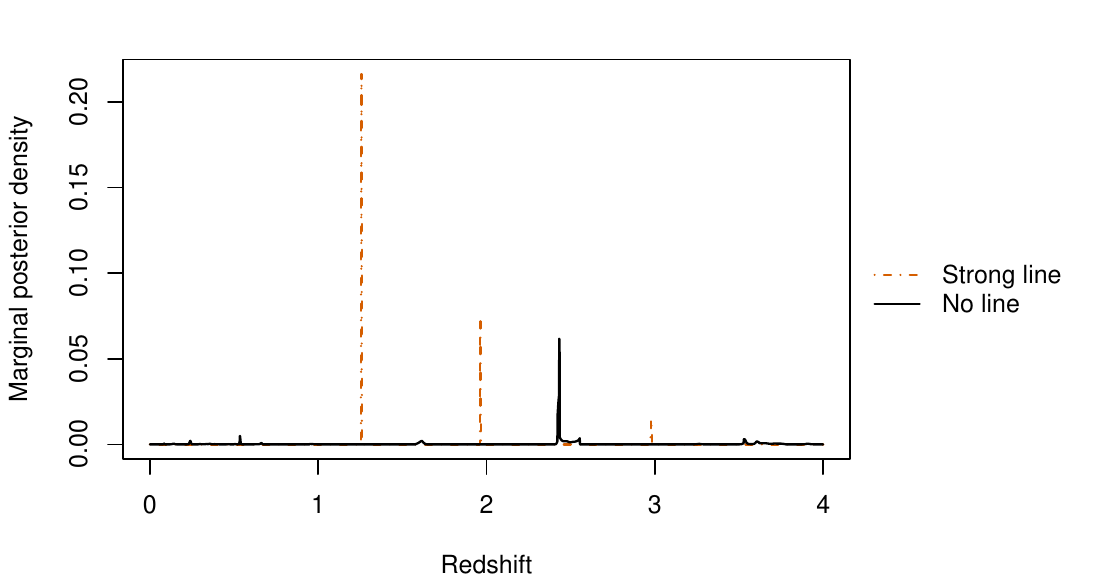}
    \caption{Discrete approximation to the posterior marginal density (grid resolution of 0.001) for two simulated datasets, one with a strong line present, and the other with no lines present. Both are normalized so that the resulting probability mass functions sum to one.}
    \label{fig:postmarg}
\end{figure}

Figure~\ref{fig:postmarg} shows a discrete approximation to the marginal posterior of redshift (described in Section~\ref{sec:modcomp}) when a strong signal is present, and when no signal is present. In the no-signal case, the estimated posterior is diffuse, placing small probabilities on many redshift values. In the strong-signal case, the estimated posterior places higher mass on a much smaller set of redshift values; however, the resulting posterior is still multimodal, with large separation between modes. This illustrates a difficult interaction between redshift and signal strength that makes jointly modeling the two especially computationally challenging. Thus, even sophisticated MCMC methods including tempering-based algorithms \citep{geyerAnnealingMarkovChain1995} or those with derivative-based proposals \citep{robertsOptimalScalingDiscrete1998}, may mix slowly and can take days to produce reliable posterior estimates. This issue is particularly relevant to the current work, as we use Bayes factors for emission line detection, which typically require substantially longer MCMC runs than standard posterior estimation to achieve stable estimates. To address these challenges, we propose an approximate approach that evaluates the posterior over a discrete grid of redshift values, leveraging conditional conjugacy to explicitly decouple the redshift and signal intensities. This parallelizable procedure is fast and scales well for massive datasets, reducing computation time from hours or days (as with MCMC) to seconds.

It is worth noting that taking a frequentist approach is difficult in this setting, since the inference problem is irregular. In particular, under the null model of no signal, redshift is not identifiable, and thus classical asymptotic results such as Wilk's theorem \citep{wilksLargeSampleDistributionLikelihood1938} do not apply. To address this, early work by  \citet{daviesHypothesisTestingWhen1977, daviesHypothesisTestingWhen1987} introduced a theoretical framework for hypothesis testing based on a likelihood ratio process indexed by the non-identifiable parameter. This line of work was later extended in the high-energy physics literature to address the  \emph{look-elsewhere effect} (LEE) -- the inflation of false discovery probability that arises when searching for a signal over a region, rather than testing at a prespecified location \citep{grossTrialFactorsLook2010}. However, existing frequentist methods that handle the LEE are not directly applicable to the problem at hand as the required regularity conditions \citep[see, e.g.,][]{algeriTestingOneHypothesis2021} are not met. For example, for each fixed value of redshift, the number of potentially active emission lines (and thus the number of constraints) is changing. This causes the limiting process to change along with the value of redshift, which is a difficulty not accounted for in existing theoretical work. Furthermore, resampling-based approaches \citep{hansenInferenceWhenNuisance1996} that may avoid such issues are computationally infeasible in surveys containing millions of spectra and a desired significance level of $3\sigma$ (i.e., level $\Pr(Z > 3) \approx 0.00135$ in the one-sided case, where $Z \sim N(0, 1)$). 

An alternative approach frames the problem of detecting a signal with unknown location as a multiple testing problem, despite the search space being inherently continuous. This can be addressed by discretizing the space and applying, for instance, a Bonferroni correction. However, this will be extremely conservative when the search space is large \citep{algeriMethodsCorrectingLookelsewhere2016} as is the case in the current application, where thousands of tests would need to be conducted. 

Finally, \citet{bayerLookelsewhereEffectUnified2020} attempt to address the LEE by using Bayes factors, but they employ improper priors, which lead to non-unique Bayes factors. The use of improper priors with Bayes factors requires care to avoid introducing arbitrary constant terms that can lead to unreliable inference \citep{bergerIntrinsicBayesFactor1996}.

This is the first work to allow for simultaneous detection of multiple emission lines without reliance on templates, balancing the trade-offs necessary to be both (i) flexible enough to account for common systematic uncertainties in the model's background and noise components, and (ii) computationally fast so that the method is practical for massive data surveys even in the presence of a challenging posterior distribution.

\section{Data and physical properties} 

\begin{figure}[t]
    \centering
    \includegraphics[width=\linewidth]{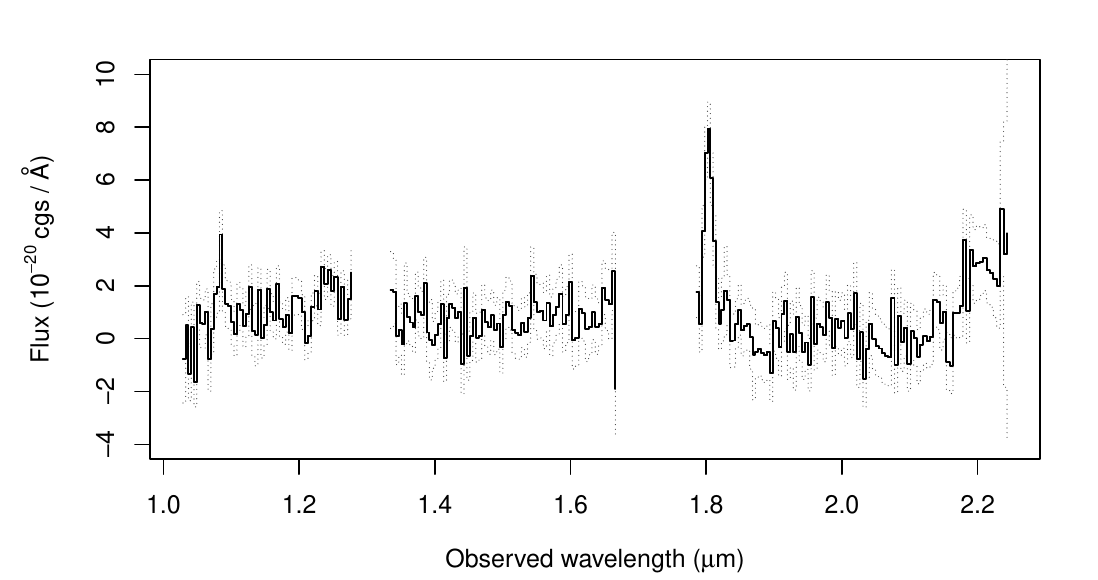}
    \caption{An observed galaxy spectrum from JWST. The data are binned corresponding to the flat rectangular nature of the plot. The grey dotted lines represent $+/-$ two standard deviations around the observed flux values using the reported measurement errors from JWST.}
    \label{fig:spec}
\end{figure}

\begin{figure}[t]
    \centering
    \includegraphics[width=\linewidth]{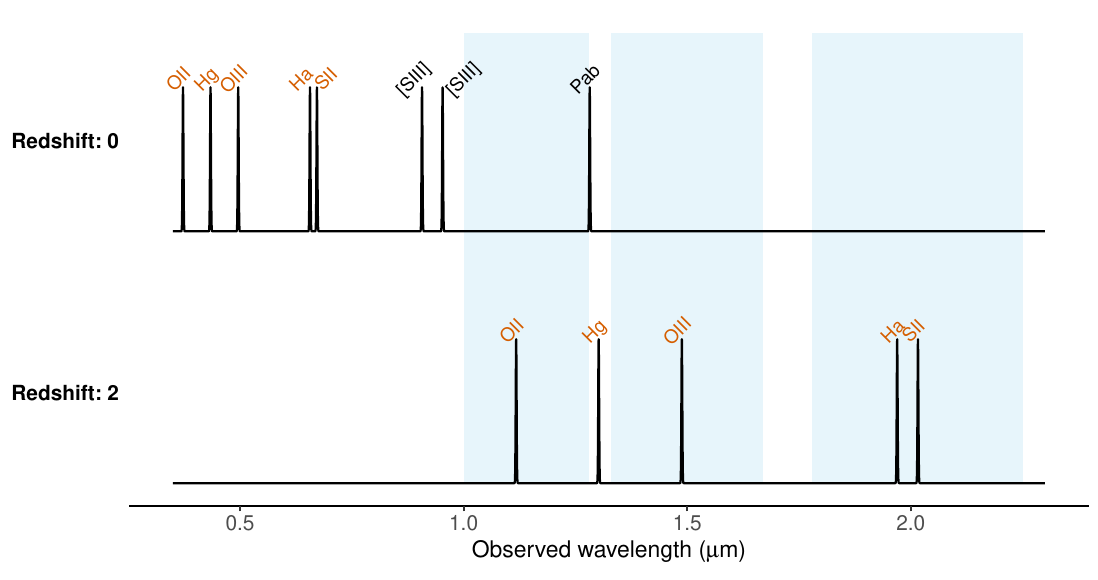}
    \caption{A comparison of observed emission line locations at redshift 0 (top) and redshift 2 (bottom). The light blue rectangles represent the observable range of the JWST detector. The OII, OIII, and SII doublets are visualized as single lines for simplicity.}
    \label{fig:redshift_lines}
\end{figure}

\subsection{Data description}
\label{sec:data}
We analyze slitless spectroscopic data obtained with JWST's NIRISS and NIRCam instruments. Because slitless spectroscopy disperses light from many galaxies simultaneously, each detector pixel records photons that may originate from multiple galaxies. Data-reduction pipelines then separate these overlapping contributions to produce a one-dimensional spectrum for each galaxy. Since each pixel accumulates a large number of photon counts (e.g., at least 500) and the observations undergo extensive calibration and processing in JWST's data-reduction pipeline, the resulting spectra are reported as real-valued \emph{flux} measurements. These are continuous-valued measurements of light intensity, accompanied by associated measurement uncertainties.

Due to the detector resolution, the observed wavelength range, $[1.03\,\mu\mathrm{m},2.25\,\mu\mathrm{m}]$ (microns), is divided into $N$ bins of similar but unequal widths. Additionally, JWST spectral data contain two detector gaps as a direct by-product of the observational strategy. More specifically, in all of the JWST spectra under consideration, observations are missing for wavelengths $[1.28\,\mu\mathrm{m}, 1.33\,\mu\mathrm{m}]$ and $[1.67\,\mu\mathrm{m},1.78\,\mu\mathrm{m}]$. Denote by $\mathcal{X}_{\text{obs}}$ the collection of intervals of wavelengths that were observed. Then, for JWST, $\mathcal{X}_{\text{obs}} = [1.03\,\mu\mathrm{m}, 2.25\,\mu\mathrm{m}] \setminus \mathcal{X}_{\text{miss}}$, where $\mathcal{X}_{\text{miss}} = [1.28\,\mu\mathrm{m}, 1.33\,\mu\mathrm{m}]\cup[1.67\,\mu\mathrm{m}, 1.78\,\mu\mathrm{m}]$.  

In our analysis, we consider 213 high-quality JWST spectra that experts have flagged as containing emission lines. These spectra have accompanying redshift estimates determined by astronomers using a procedure described in \citet{hubertyNIRISSPASSAGESpectroscopic2026}. Briefly, an automated line-fitting procedure is first applied, assuming that the strongest emission line corresponds to the element H$\alpha$. Experts then visually assess the resulting fit and test alternative emission line identifications until a satisfactory fit is obtained. An example of a high-quality JWST spectrum is shown in Figure~\ref{fig:spec}. In addition, we consider 2800 JWST spectra that were not flagged as high quality and were not selected by experts as containing emission lines. These spectra were not accompanied by astronomer-provided redshift estimates in the dataset used in this study, so the analysis in Section~\ref{sec:results} provides novel redshift estimates for these galaxies.

\subsection{Emission lines and redshift}
\label{sec:lines}
We are interested in the possible detection of 13 emission lines from various elements, each with a known lab-frame wavelength (in $\mu$m): OII (0.37271, 0.37299), H$\gamma$ (0.434), H$\beta$ (0.486), OIII (0.496, 0.501), H$\alpha$ (0.656), NII (0.658), SII (0.672, 0.673), [SIII] (0.907, 0.953), and Pa$\beta$ (1.282). Some elements produce multiple closely spaced lines, known as \emph{doublets}, corresponding to the pairs of wavelengths listed above. The intensities of lines within each doublet are related by known constant factors, following well-established physical relationships (see Section~\ref{sec:line_eligibility}). Similarly, the hydrogen lines (H$\gamma$, H$\beta$, and H$\alpha$) have a known ordering in their relative strengths, with the exact values depending on the amount of dust obscuration present in each galaxy (see Section~\ref{sec:model_specific_priors}).  

When a galaxy is observed, each lab-frame wavelength, $x_k^{\text{lab}}$, is shifted by the unknown redshift, $\zeta$, so that the observed wavelength is given by $x_k^{\text{lab}}(1 + \zeta)$. This deterministic scaling determines which emission lines fall within the observable wavelength region, $\mathcal{X}_{\text{obs}}$, linking observed emission line location to galaxy redshift. 

Figure~\ref{fig:redshift_lines} illustrates how the emission lines move into and out of the observable range as a function of redshift. This highlights why simultaneous detection of multiple lines is necessary: some lines may be unobservable at certain redshifts, while others remain detectable, providing complementary information. 

\section{Model likelihood} 
\label{sec:likelihood}
Assume the flux values, denoted $Y_i$ for $i = 1,...,N$, are normally distributed. This section describes how the likelihood can be specified to incorporate not only the expected structure of the flux values but also the rules governing emission-line eligibility, physical constraints, and the covariance structure of the observations.

\subsection{Mean structure}
Each observation corresponds to the central wavelength of the $i$-th bin, denoted $x_i$. The conditional mean of each $Y_i$ is modeled as
\begin{equation}
    \text{E}[Y_i \mid \alpha, f_b, \eta, \zeta, \delta, x_i] = \alpha + f_b(x_i) + \sum_{k=1}^{13} \eta_k(\zeta) s_k(x_i, \zeta, \delta),
\end{equation} 
where $\alpha$ is an intercept term, $f_b(\cdot)$ is an unknown background function, and the sum represents the signal contribution from all 13 emission lines of interest. 

The background function, $f_b(\cdot)$, is expected to vary smoothly over the wavelength range, although little additional structure is known \emph{a priori}. The limited knowledge about $f_b(\cdot)$ arises from the data processing required to produce the spectra and from possible contaminants (``glitches'') in the data. The background is modeled as
\begin{equation*}
    f_b(x) = \sum_{j=1}^J \beta_j b_j(x),
\end{equation*}
where $b_j(\cdot)$ denotes the $j$-th cubic B-spline basis function. We fix $J = 15$ for the analysis in Section~\ref{sec:results}. Sensitivity to this choice is examined in Section~\ref{app:bg} of the Supplementary Material and summarized in Section~\ref{sec:final_remarks}.

The signal component is a linear combination of emission line intensities, $\eta_k(\zeta)$, and known line-shape functions, $s_k(\cdot, \zeta, \delta)$, which depend on wavelength, the redshift parameter, $\zeta$, and the line-width parameter, $\delta$. Each line-shape function is modeled as a Gaussian density centered at the redshifted lab-frame wavelength and evaluated over the observed wavelength range $x \in \mathcal{X}_{\text{obs}}$:
\begin{equation}
\label{eq:lineshape}
    s_{k}(x, \zeta, \delta) = \frac{1}{\sqrt{2\pi \delta^2w^2}}\exp\left\{-\frac{1}{2\delta^2 w^2}[x - x^{\text{lab}}_{k}(1 + \zeta)]^2\right\},\; k = 1,...,K,
\end{equation}
where $w$ is the median pixel width. The line-width parameter, $\delta$, scales the width of each emission line width in units of $w$.

\subsection{Emission line eligibility and constraints}
\label{sec:line_eligibility}
Let $\eta_k(\zeta)$ denote the signal intensity corresponding to the $k$-th emission line at redshift $\zeta$, and define the column vector of all 13 signal intensities as 
\begin{equation*}
    \eta(\zeta) = (\eta_1(\zeta),
    \dots, \eta_{13}(\zeta))'
\end{equation*}
where the prime symbol indicates the transpose. A line is considered \emph{eligible} at redshift $\zeta$ if its redshifted lab-frame wavelength falls in the observable region. Hence, the set of eligible indices is 
\begin{equation*}
    \mathcal{E}(\zeta) = \{k \; : \; x_k^{\text{lab}}(1+\zeta) \in \mathcal{X}_{\text{obs}}\}.
\end{equation*}
For $k \not \in \mathcal{E}(\zeta),$ we set $\eta_k(\zeta) = 0$.

Certain pairs of emission line intensities satisfy deterministic physical relationships, as described in Section~\ref{sec:lines}. Let $\mathcal{C}$ denote the set of such ordered pairs $(k, \ell)$ with $k < \ell$. Whenever both lines in a constrained pair are eligible, the corresponding signal intensities satisfy 
\begin{equation*}
    \eta_{\ell}(\zeta) = c_{k\ell}\eta_k(\zeta), \quad (k, \ell)\in\mathcal{C}; \; k, \ell \in \mathcal{E}(\zeta).
\end{equation*}
Requiring both lines to be eligible in order to impose the constraint is necessary here to avoid the case where one line falls in the observable region, but the other does not. In that case, we still need to estimate the eligible line freely. The specific emission line constraints are 
\begin{equation}
\begin{aligned}
    \eta_2(\zeta) &= \eta_1(\zeta), \quad
    \eta_6(\zeta) = 0.336\,\eta_5(\zeta), \\
    \eta_{10}(\zeta) &= 0.738\,\eta_9(\zeta), \quad
    \eta_{12}(\zeta) = 0.4\,\eta_{11}(\zeta).
\end{aligned}
\end{equation}

To avoid estimating redundant parameters, define $\mathcal{K}(\zeta) \subseteq \mathcal{E}(\zeta)$ as the set of eligible indices that remain after removing the second index of each constrained pair when both members are eligible (so the smaller index, $k$, is retained by convention). Formally,
\begin{equation}
\label{eq:K_set}
    \mathcal{K}(\zeta) = \mathcal{E}(\zeta) \setminus \{\ell \; : \; (k, \ell) \in \mathcal{C}, \text{ and } k, \ell \in \mathcal{E}(\zeta)\}.
\end{equation}
The reduced signal intensity vector can now be defined as the subvector 
\begin{equation*}
    \eta_{\text{sub}}(\zeta) = \{\eta_k(\zeta)\}_{k \in \mathcal{K}(\zeta)},
\end{equation*}
and we only need to estimate $\eta_{\text{sub}}(\zeta)$.

To express the mean structure in terms of $\eta_{\text{sub}}(\zeta)$, begin by defining the full signal matrix at $(\zeta, \delta)$ as the matrix with elements: 
\begin{equation}
    [W(\zeta, \delta)]_{ik} := s_k(x_i, \zeta, \delta), \quad i = 1,\dots, N, \; k = 1,\dots,13.
\end{equation}
Columns corresponding to deterministically related lines are combined through linear transformations so that only one coefficient in each constrained pair is estimated (again, using the smaller index by convention). For example, if both $\eta_5$ and $\eta_6$ are eligible, their respective columns in $W(\zeta, \delta)$ are merged into a single column with entries
\begin{equation*}
    s_5(x_i, \zeta, \delta) + 0.336 s_6(x_i, \zeta, \delta), \quad i = 1,\dots,N.
\end{equation*}
Columns corresponding to ineligible lines are then removed, yielding the reduced signal matrix, $W_{\text{sub}}(\zeta, \delta)$, of dimension $N \times |\mathcal{K}(\zeta)|$, where $|\mathcal{K}(\zeta)|$ denotes the cardinality of $\mathcal{K}(\zeta)$.

Note that while removing columns from $W(\zeta, \delta)$ and entries from $\eta(\zeta)$ corresponding to ineligible lines is not strictly necessary, reducing the dimension to obtain $W_{\text{sub}}(\zeta, \delta)$ leads to straightforward derivations for the posterior distribution in terms of generalized least-squares estimates (see Sections~\ref{app:deriv} and~\ref{app:pointests} of the Supplementary Material).

\subsection{Covariance structure}
The covariance of $Y = (Y_1,\dots,Y_N)'$ is modeled as
\begin{equation}
\label{eq:cov}
    S(\sigma^2) = \sigma^2I_{N} + C,
\end{equation}
where $I_{N}$ is the $N\times N$ identity matrix, $\sigma^2$ is an unknown scalar, and $C$ is a known diagonal matrix with entries $C_{ii} = c_i^2$ given by the reported measurement errors. In practice, the matrix $C$ is usually diagonal, but the methodology remains unchanged if it were not.

The additive structure in equation~\eqref{eq:cov} allows explicit incorporation of measurement error information from JWST through the matrix $C$, which is necessary to prevent false detections. Measurement errors can become large near the edges of the observable region, $\mathcal{X}_{\text{obs}}$, producing extreme flux values that might otherwise be mistaken for emission lines. The homoskedastic noise term, $\sigma^2 I_N$, is included to capture additional statistical variability. 

\subsection{Linear model formulation}
We can now express the model for the flux as a linear model for fixed redshift and line-width pairs, $(\zeta, \delta)$, as 
\begin{equation}
\label{eq:model}
    Y = \alpha 1_N + X\beta + W_{\text{sub}}(\zeta, \delta) \eta_{\text{sub}}(\zeta) + \varepsilon,
\end{equation}
where $1_N$ is the length-$N$ vector of ones, $X$ is a matrix of elements $X_{ij} = b_j(x_i)$ for $i = 1,\dots,N$ and $j = 1,\dots,J$, and $\beta = (\beta_1,\dots,\beta_J)'$. Additionally, assume $\varepsilon = (\varepsilon_1,\dots, \varepsilon_N)' \sim N(0, S(\sigma^2))$. To simplify posterior derivations, we use orthogonalized versions of $X$ and $W_{\text{sub}}(\zeta, \delta)$ with respect to $S(\sigma^2)$ defined in Section~\ref{app:orthog} of the Supplementary Material.  

\section{Priors}
\label{sec:prior}

We consider two competing models for the observed spectrum. Under the null model, $M_0$, no emission lines are present. Under the alternative model, $M_1$, one or more emission lines are present, with unknown redshift $\zeta$, line width $\delta$, and signal intensities $\eta_{\text{sub}}(\zeta)$. The background parameters $\alpha$ and $\beta$, as well as the noise variance $\sigma^2$, are common to both models. In this section, we specify prior distributions for all parameters appearing under each model.

For the full parameter vector, $\theta = (\alpha, \beta, \sigma^2, \eta(\zeta), \zeta, \delta)',$ we assume the following prior structure
\begin{equation}
\label{eq:priorstr}
    p(\theta) \propto p(\alpha)p(\beta)p(\sigma^2)p(\eta_{\text{sub}}(\zeta) \mid \zeta) p(\zeta)p(\delta),
\end{equation}
where $p(\cdot)$ denotes a probability density function with respect to the appropriate dominating measure. Here, $\eta_{\text{sub}}(\zeta)$ is defined as in Section~\ref{sec:likelihood} and contains the free elements of $\eta(\zeta)$ at redshift $\zeta$. The remaining elements in $\eta(\zeta)$ are deterministically specified by the redshift and $\eta_{\text{sub}}(\zeta)$. Therefore, they affect the prior only by a constant factor. Priors for the parameters contributing to the product on the right-hand side of equation~\eqref{eq:priorstr} are specified in the following subsections, along with hyperparameter choices motivated by existing astrophysical theory and expert input. A sensitivity analysis is presented in Section~\ref{app:sens} of the Supplementary Material and summarized in Section~\ref{sec:final_remarks}.

\subsection{Priors common to both models}
For the intercept term, we assume $\alpha \sim N(a_0, b_0^2)$ with $a_0 = 5$ and $b_0 = 10$, reflecting the prior belief that flux values are typically nonnegative, while allowing for negative reported flux values arising from the JWST pre-processing pipeline.

For the B-spline background coefficients, we assume $\beta_j \sim N(a_1, b_1^2),$ with $a_1 = 0$ and $b_1 = 30$, for $j = 1,\dots, 15$. The prior mean of zero reflects the initial best guess of a constant background,
while the large prior variance and relatively large number of basis functions ($J = 15$) allow flexibility to capture contaminant and spectral overlap effects. 

We fix the noise variance, $\sigma^2$, at its maximum-likelihood estimate under $M_1$, denoted $\hat{\sigma}^2$ (see Section~\ref{app:var_est} of the Supplementary Material), and use this fixed value in both $M_0$ and $M_1$ for all subsequent inference. Although a conjugate inverse-Gamma prior on $\sigma^2$ would be computationally convenient, it would require inducing an explicit dependence of the prior variances for $\alpha, \beta$, and $\eta_{\text{sub}}(\zeta)$ on $\sigma^2$, which is not physically meaningful in this setting. Non-conjugate alternatives that integrate over $\sigma^2$ avoid this dependence, but are computationally costly since $\sigma^2$ enters the covariance structure and would require repeated inversions of relatively large matrices.

\subsection{Model specific priors}
\label{sec:model_specific_priors}
We model the absence of a signal (the null model) via a point mass prior at $\eta(\zeta) \equiv 0$. Since no emission lines are present under this model, no prior for the redshift, $\zeta$, is required.

Under the alternative model, in which at least one line is present, we place a scaled Beta(3, 3) prior on redshift, $\zeta$, over the interval $(0, 4]$, yielding a symmetric, parabolic density peaked at $\zeta = 2$. This reflects a weak prior preference for the OII and H$\alpha$ lines above, since the prior mean is set near their eligible range. 

The line-width parameter, $\delta$, is given a uniform prior over $[1, 3]$, reflecting the expected range of narrow emission line widths relative to the data resolution (recall that the Gaussian line variance is $\delta^2 w^2$, where $w$ is the median pixel width). In practice, we discretize both $\zeta$ and $\delta$ to enable an efficient computation strategy described in Section~\ref{sec:modcomp} below. 

The signal intensities, conditional on $\zeta$, are given the following product prior:
\begin{equation*}
    p(\eta_{\text{sub}}(\zeta) \mid \zeta) 
    \propto 1_{\{\eta_{\text{sub}}(\zeta) \in \mathcal{A}(\zeta)\}}\prod_{k \in \mathcal{K}(\zeta)}p(\eta_k(\zeta)),
\end{equation*}
where $p(\eta_k(\zeta))$ denotes a density with support on $[0, \infty)$, and we define \begin{equation}
\label{eq:cons}
    \mathcal{A}(\zeta) := \left\{\eta_{\text{sub}}(\zeta) : \frac{\eta_4(\zeta)}{\eta_3(\zeta)} \geq 2.174, \; \frac{\eta_7(\zeta)}{\eta_4(\zeta)} \geq 2.86, \; \eta_k(\zeta) \geq 0, \;k \in \mathcal{K}(\zeta)\right\}.
\end{equation}
The set $\mathcal{A}(\zeta)$ encodes the physical inequality constraints between the hydrogen line intensities mentioned in Section~\ref{sec:lines}. Each $p(\eta_k(\zeta))$ is taken to be the density of a normal distribution with parameters $a_{2,k}$ and $b_{2,k}$, leading to the multivariate truncated normal density:
\begin{equation}
\label{eq:tnorm}
    p(\eta_{\text{sub}}(\zeta) \mid \zeta) \propto 1_{\{\eta_{\text{sub}}(\zeta) \in \mathcal{A}(\zeta)\}} \prod_{k \in \mathcal{K}(\zeta)}\exp\left\{-\frac{1}{2b_{2,k}^2}(\eta_k(\zeta) - a_{2,k})^2\right\},
\end{equation} 
where the hyperparameter vectors
\begin{equation}
\label{eq:eta_hyperparams}
    a_2 = (a_{2,1},\dots, a_{2,13})', \quad b_2 = (b_{2,1},\dots, b_{2,13})'
\end{equation}
collect the hyperparameters for all 13 emission lines. 

The multivariate truncated normal prior is convenient because it directly enforces linear inequality constraints among the hydrogen emission line intensities, while preserving some analytic tractability, thereby minimizing computational costs (see Section~\ref{sec:modcomp}). Other prior families require constrained optimization or specialized sampling, which can substantially increase computation time.

The specific values of $a_2$ and $b_2$ used for JWST are
\begin{equation*}
a_2 = (0.02, 0.02, 0.0092, 0.02, 0.2, 0.596, 0.2, 0.02, 0.02, 0.0271, 0.02, 0.05, 0.02)',
\end{equation*}
and
\begin{equation*}
    b_2 = (0.01, 0.01, 0.01, 0.01, 1, 1, 1, 0.01, 0.01, 0.01, 0.01, 0.01, 0.01)'.
\end{equation*}
For most emission lines, the prior mean is quite small, with inequality constraints in \eqref{eq:cons} imposed as equality constraints on the appropriate entries of $a_2$. Additionally, when strong emission lines are present, they are more likely to be either OII or H$\alpha$. This is reflected in the prior means, two orders of magnitude larger than the rest. 

Finally, for JWST, the model priors are $p(M_0) = 0.9$ and $p(M_1) = 0.1$, reflecting prior belief that galaxies without observable emission lines are much more common than those with observable emission lines.

\section{Model comparison and estimation}
\label{sec:modcomp}
For signal detection, we compare the no-signal model, $M_0$, with the signal model, $M_1$, via their posterior odds,
\begin{equation}
\label{eq:post_odds}
    \frac{p(M_1 \mid y)}{p(M_0 \mid y)} 
    = \frac{p(y \mid \hat{\sigma}^2, M_1)}{p(y \mid \hat{\sigma}^2, M_0)} \times \frac{p(M_1)}{p(M_0)},
\end{equation}
where $p(y \mid \hat{\sigma}^2, M_h)$ is the marginal density of the observed data, $y$, under model $M_h, \; h = 0,1$. The first term in the product on the right-hand side of \eqref{eq:post_odds} is the Bayes factor, $\text{BF}_{10}$, and the second term is the prior odds. The no-signal model, $M_0$, includes the intercept and background parameters, $\alpha$ and $\beta$, as well as the variance $\sigma^2$. The signal model, $M_1$, additionally includes the free signal intensities, $\eta_{\text{sub}}(\zeta)$, the redshift, $\zeta$, and the line-width parameter, $\delta$.

\subsection{Marginal density under the null model}
To compute the marginal density under $M_0$, recall that $Y = \alpha1_N + X\beta + \varepsilon$, where $\alpha \sim N(a_0, b_0^2), \beta \sim N(0, b_1^2 I),$ and $\varepsilon \sim N(0, S(\sigma^2))$. Because $Y \mid \hat{\sigma}^2, M_0$ is a linear combination of normally distributed random variables, it follows that
\begin{equation*}
    Y \mid \hat{\sigma}^2, M_0 \sim N(a_0 1_N, S(\hat{\sigma}^2) + b_0^21_N 1_N' + b_1^2XX'),
\end{equation*}
where $S(\hat{\sigma}^2)$ is defined as in equation~\eqref{eq:cov} with $\hat{\sigma}^2$ in place of $\sigma^2$. 

\subsection{Marginal density under the alternative model}
Under $M_1$, in which at least one emission line is present, the goal is to compute $p(y \mid \hat{\sigma}^2, M_1)$ based on the prior on $\eta_{\text{sub}}(\zeta) \mid \zeta$ in equation~\eqref{eq:tnorm}. Directly integrating over $\eta_{\text{sub}}(\zeta), \alpha, \beta, \zeta,$ and $\delta$ is intractable, so we consider a fine grid of $(\zeta, \delta)$ pairs and proceed as follows. Let $\zeta_1,\dots,\zeta_R$ denote a predefined, equally spaced grid over $(0, 4]$, chosen to match the desired redshift resolution. Similarly, let $\delta_1,\dots,\delta_D$ denote a predefined grid over $[1, 3]$. All calculations for $M_1$ are performed over the $R \times D$ grid of $(\zeta_r, \delta_\ell)$ pairs.

\subsubsection{\texorpdfstring{Integration over $\eta_{\text{sub}}(\zeta), \alpha$, and $\beta$}{Integration over eta\_sub(zeta), alpha, and beta}}
\label{sec:integration}
Consider a fixed $(\zeta_r, \delta_\ell)$ pair. Rather than working with the truncated multivariate normal prior on $\eta_{\text{sub}}(\zeta_r) \mid \zeta_r$ in equation~\eqref{eq:tnorm} directly, we introduce an \emph{encompassing model} in which the truncated prior is replaced by its unconstrained multivariate normal counterpart. Specifically, we define
\begin{equation}
\label{eq:encomp}
    \eta^*_{\text{sub}}(\zeta_r) \mid \zeta_r \sim N(a_2(\zeta_r), B_2(\zeta_r)),
\end{equation}
where $\eta^*_{\text{sub}}(\zeta_r)$ denotes the unconstrained version of $\eta_{\text{sub}}(\zeta_r)$, and 
\begin{equation}
\label{eq:hyperparams}
    a_2(\zeta_r) = \{a_{2,k}\}_{k \in \mathcal{K}(\zeta_r)}, \quad B_2(\zeta_r) = \text{diag}(\{b_{2,k}^2\}_{k \in \mathcal{K}(\zeta_r)})
\end{equation}
denote the corresponding subvector and diagonal submatrix of hyperparameters defined in equation~\eqref{eq:eta_hyperparams} for lines with indices in $\mathcal{K}(\zeta_r)$.   

Let $p^*(y \mid \hat{\sigma}^2, \zeta_r, \delta_\ell, M_1)$ denote the density obtained by integrating out $\alpha, \beta$, and $\eta^*_{\text{sub}}(\zeta_r)$. More explicitly, this density is given by
\begin{equation}
\label{eq:pstar_density}
    \begin{aligned}
        p^*(y \mid \hat{\sigma}^2, \zeta_r, \delta_\ell, M_1) 
        &= \int p(y \mid \hat{\sigma}^2, \alpha, \beta, \eta_{\text{sub}}^*(\zeta_r), \zeta_r, \delta_\ell)\\
        &\quad\quad\quad \times p(\alpha)p(\beta)p(\eta_{\text{sub}}^*(\zeta_r) \mid \zeta_r)\,d\alpha\,d\beta\,d\eta_{\text{sub}}^*(\zeta_r),
    \end{aligned}
\end{equation}
where $p(y \mid \hat{\sigma}^2, \alpha, \beta, \eta_{\text{sub}}^*(\zeta_r), \zeta_r, \delta_\ell)$ is the conditional density of $y$ defined by equation~\eqref{eq:model} (see Section~\ref{app:orthog} of the Supplementary Material for an explicit expression). As in the null model, the conjugate Gaussian priors imply that $p^*(y \mid \hat{\sigma}^2, \zeta_r, \delta_\ell, M_1)$ is the density of a multivariate normal with mean $a_01_N + W_{\text{sub}}(\zeta_r, \delta_\ell)a_2(\zeta_r)$ and covariance matrix 
\begin{equation*}
    b_0^21_N1_N' + b_1^2XX' + W_{\text{sub}}(\zeta_r, \delta_\ell)B_2(\zeta_r)W_{\text{sub}}(\zeta_r, \delta_\ell)' + S(\hat{\sigma}^2).
\end{equation*}
A routine calculation (see Section~\ref{app:marg_deriv} of the Supplementary Material) leads to
\begin{equation}
\label{eq:ev1}
    p(y\mid \hat{\sigma}^2, \zeta_r, \delta_\ell, M_1) = p^*(y \mid \hat{\sigma}^2, \zeta_r, \delta_\ell, M_1) \frac{\Pr(\eta^*_{\text{sub}}(\zeta_r) \in \mathcal{A}(\zeta_r) \mid \hat{\sigma}^2, \zeta_r, \delta_\ell, y, M_1)}{\Pr(\eta^*_{\text{sub}}(\zeta_r) \in \mathcal{A}(\zeta_r) \mid \zeta_r, M_1)},
\end{equation}
where $\Pr(\eta^*_{\text{sub}}(\zeta_r) \in \mathcal{A}(\zeta_r) \mid \zeta_r, M_1)$ and $\Pr(\eta^*_{\text{sub}}(\zeta_r) \in \mathcal{A}(\zeta_r) \mid \hat{\sigma}^2, \zeta_r, \delta_\ell, y, M_1)$ are the prior and posterior probabilities that $\eta^*_{\text{sub}}(\zeta_r)$ satisfies the constraints in equation~\eqref{eq:cons}. Details for computing these probabilities are provided in Section~\ref{app:deriv} of the Supplementary Material; in particular, each probability reduces to a nonnegativity constraint on a known multivariate normal distribution, which can be computed efficiently.

\subsubsection{\texorpdfstring{Integration over $\delta$ and $\zeta$}{Integration over delta and zeta}}
Since $\delta$ has been discretized, integrating out $\delta$ for each fixed $\zeta_r$ is approximated by a Riemann sum over its support:
\begin{equation*}
    p(y \mid \hat{\sigma}^2, \zeta_r, M_1) = \sum_{\ell = 1}^D p(y \mid \hat{\sigma}^2, \zeta_r, \delta_\ell, M_1)p(\delta_\ell)\Delta_{\delta},
\end{equation*} 
where $\Delta_{\delta}$ denotes the grid spacing for $\delta$ (with $\Delta_{\delta} = 0.5$ in practice). The marginal density under the alternative model is then approximated by
\begin{equation}
\label{eq:addem}
    p(y \mid \hat{\sigma}^2, M_1) \approx \sum_{r=1}^R p(y \mid \hat{\sigma}^2, \zeta_r, M_1)p(\zeta_r)\Delta_{\zeta},
\end{equation}
where $\Delta_\zeta$ denotes the grid spacing for $\zeta$. For spectra observed by JWST, we use $\Delta_\zeta = 0.001$. For the chosen priors, equation~\eqref{eq:ev1} must be evaluated for all $6 \times 4000 = 24{,}000$ $(\zeta_r, \delta_\ell)$ pairs, but the computations can be easily parallelized.

\subsection{Redshift estimation under the alternative model}
The posterior marginal distribution $\zeta$ is approximated by
\begin{equation}
\label{eq:pmf}
    p(\zeta_r \mid y, \hat{\sigma}^2, M_1) \approx \frac{p(y \mid \hat{\sigma}^2, \zeta_r, M_1)p(\zeta_r)\Delta_\zeta}{\sum_{r=1}^R p(y \mid \hat{\sigma}^2, \zeta_r, M_1) p(\zeta_r)\Delta_\zeta},
\end{equation} 
for $r = 1,\dots,R$. The discrete approximation in equation~\eqref{eq:pmf} is used directly for model selection and redshift estimation, since the grid resolution is sufficiently fine for inference. 

As a point estimate of the redshift, we use the approximate marginal maximum a posteriori (MAP) estimator,
\begin{equation}
\label{eq:postmapest}
    \hat{\zeta} = \arg\max_{r}p(\zeta_r \mid y, \hat{\sigma}^2, M_1).
\end{equation}
Point estimates for the remaining parameters are taken as their posterior means, conditional on $\zeta = \hat{\zeta}$ and $\sigma^2 = \hat{\sigma}^2$. Derivations for these estimates are provided in Section~\ref{app:pointests} of the Supplementary Material.

Uncertainty in redshift estimation is summarized using approximate marginal highest posterior density (HPD) sets for $\zeta$. Because the posterior, $p(\zeta \mid \hat{\sigma}^2, y, M_1)$, is evaluated on a discrete grid and is often multimodal, the HPD region typically consists of multiple disjoint intervals rather than a single contiguous range. To construct a $(1-\alpha)$ HPD set, we order the grid points $\{\zeta_r\}_{r=1}^R$ by decreasing posterior probability and include the most probable values until the cumulative posterior mass reaches at least $1 - \alpha$. Adjacent included grid points are then grouped into contiguous components, each corresponding to a local mode of the posterior distribution. The union of these components forms the approximate $(1-\alpha)$ HPD set.

\section{Results on JWST data}
\label{sec:results}

\begin{figure}[t]
    \centering
    \includegraphics[width=\linewidth]{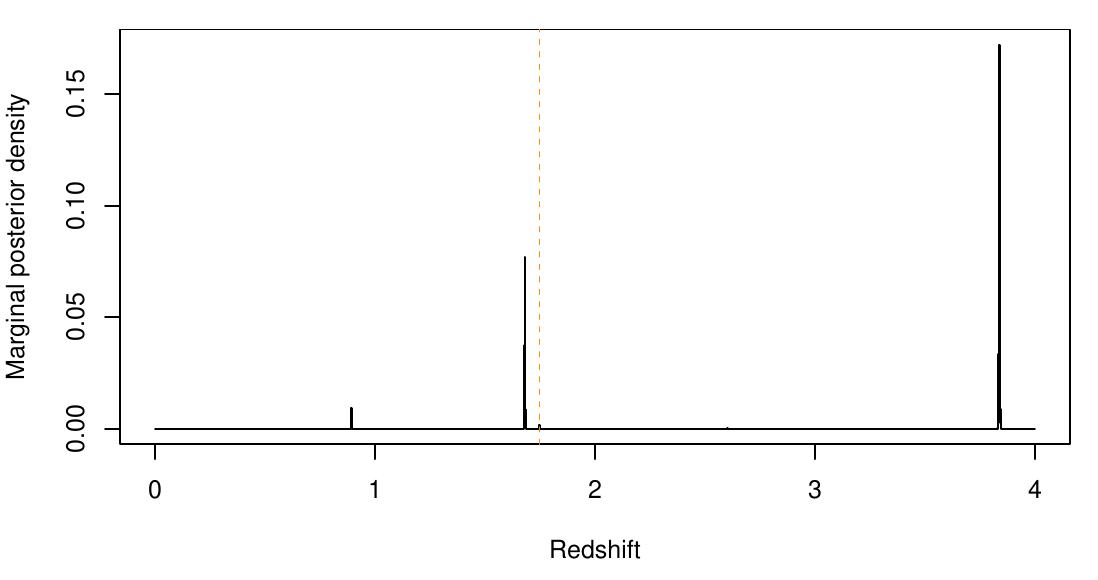}
    \caption{Discrete approximation to the posterior marginal density for spectrum 00004. The dashed orange line indicates the redshift estimate provided by astronomers.}
    \label{fig:specmargres}
\end{figure}

To assess the performance of the proposed procedure on JWST data, we begin with three individual spectra that illustrate key components of the analysis. We then present aggregate results for the 213 high-quality (labeled) spectra and compare our emission-line detection decisions and redshift estimates with those provided by JWST astronomers. Finally, we analyze 2800 lower-quality (unlabeled) JWST spectra that have not been assigned expert classifications, owing to both the difficulty of characterizing these spectra and the size of the dataset. A subset of the JWST spectra (galaxies 00004, 00033, and 02263), along with associated astronomer-provided redshift annotations and R code required to reproduce these results, is provided in the Data and R Code Supplement. The full dataset is subject to JWST data-sharing restrictions but is available upon request.

Model fit assessment and a sensitivity analysis of prior hyperparameter choices, including an investigation of false positive rates, are presented in Sections~\ref{app:model_check} and~\ref{app:sens} of the Supplementary Material. Emission line detection is based on the astronomers' prior model probabilities, $p(M_0) = 0.9$ and $p(M_1) = 0.1$, given in Section~\ref{sec:prior}. We select $M_1$ if the log posterior odds are greater than zero, and $M_0$ otherwise.

\subsection{Analyzing individual spectra}
We analyze spectra from three illustrative galaxies in the high-quality (labeled) JWST sample, corresponding to galaxy IDs 00004, 00033, and 02263 in the Data and R Code Supplement. Spectrum 00004 has a relatively constant background and a single prominent Gaussian line in the observed range. Emission line detection is therefore straightforward, but redshift estimation is more difficult, as the marginal posterior of redshift is multimodal. Spectrum 00033 has a more complex background. However, multiple emission lines are detected across the observed range, which more tightly constrain the redshift. In this case, the marginal posterior is unimodal. Finally, spectrum 02263 has a prominent, non-constant background and relatively weak emission lines, making emission lines difficult to identify. The evidence for detection is weak (although emission lines are detected), and the redshift is only weakly constrained by the observed data. Together, these three cases demonstrate how background structure and the number of detectable emission lines influence both the evidence for emission line detection and the shape of the redshift posterior. While these examples highlight recurring features in the data, the full sample exhibits a broader range of background shapes and emission line configurations. 

\subsubsection{Galaxy 00004}
\begin{figure}[t]
    \centering \includegraphics[width=\linewidth]{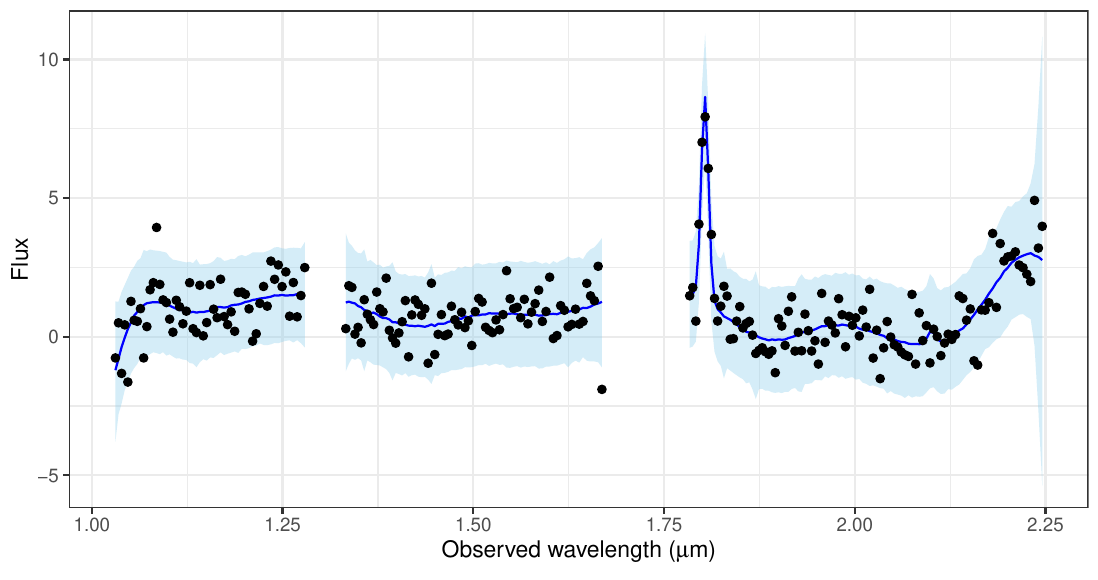}
    \caption{A posterior predictive plot assessing model fit for spectrum 00004 in Figure~\ref{fig:spec}. The black dots show the observed flux values per wavelength. Here, 5000 replicate datasets are drawn from the posterior predictive distribution. The pointwise median per wavelength is shown in dark blue, and the pointwise 2.5\% and 97.5\% pointwise quantiles as the lower and upper bounds of the light blue region, respectively.}
    \label{fig:post_pred_01}
\end{figure}

Consider the JWST spectrum shown in Figure~\ref{fig:spec}. The log posterior odds for $M_1$ versus $M_0$ is 26.47, corresponding to a posterior probability of $M_1$ being effectively 1. For reference, the corresponding log Bayes factor is 28.67. These values provide extremely strong evidence for the presence of at least one emission line, consistent with astronomers' high confidence that this spectrum contains emission lines.

Based on the strong evidence for detection, we proceed to estimate the galaxy's redshift. Figure~\ref{fig:specmargres} displays the discrete approximation to the posterior marginal distribution for redshift. The posterior is multimodal, with four modes (the fourth being quite small). The dominant mode is centered at 3.837, the marginal MAP for redshift. The corresponding marginal 99.865\% HPD set (the $3\sigma$ set, reflecting astrophysical convention) is the union $[0.890, 0.894]\cup[1.678, 1.684]\cup[1.744, 1.749]\cup[3.831, 3.844]$. This set contains the redshift value 1.74734 reported by astronomers, although the posterior marginal MAP differs substantially from their point estimate. Importantly, the astronomers' redshift is provided without an accompanying measure of uncertainty and should not be regarded as the ground truth. In this case, the overlap between the HPD set and the astronomers' estimate offers partial agreement, while the multimodality of the posterior highlights substantial uncertainty. 

\begin{figure}[ht]
    \centering
    \includegraphics[width=\linewidth]{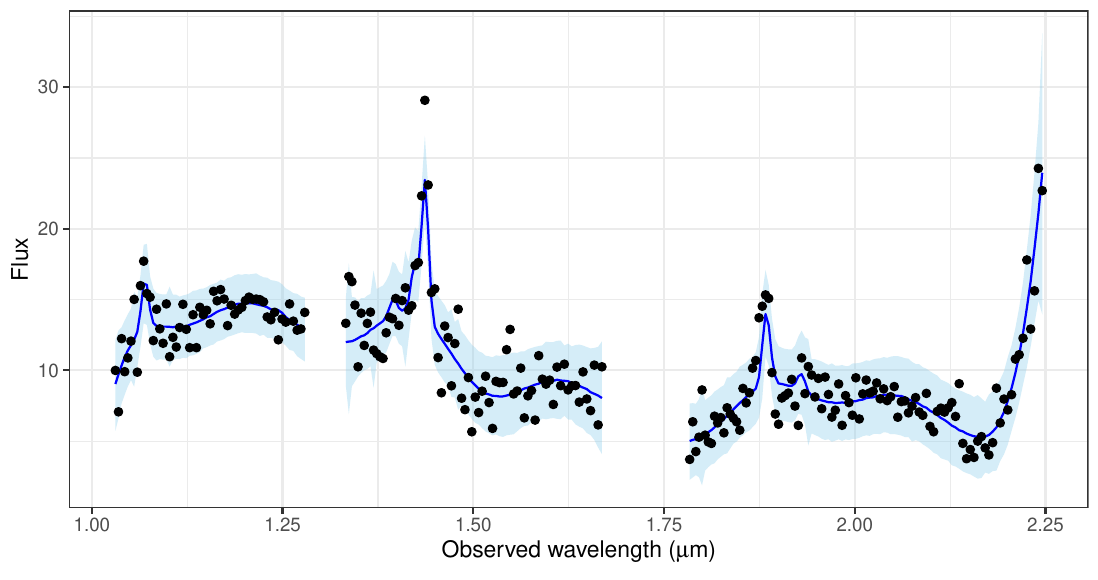}
    \caption{A posterior predictive plot assessing model fit for spectrum 00033. The black dots show the observed flux values per wavelength. Here, 5000 replicate datasets from the posterior predictive distribution are drawn, with the pointwise median per wavelength shown in dark blue, and the pointwise 2.5\% and 97.5\% pointwise quantiles as the lower and upper bounds of the light blue region, respectively.}
    \label{fig:spec_04}
\end{figure}

Next, we report the set of non-zero emission lines at the marginal MAP redshift, and compare it to the emission lines implied by the astronomers' point estimate. At redshift 3.837, the active lines include both OII lines and H$\gamma$. In this case, the large spike in Figure~\ref{fig:spec} corresponds to the OII doublet. The posterior means of the emission line intensities (conditional on $\hat{\sigma}^2$ and $\hat{\zeta}$) are 0.044, 0.044, and 0.007, respectively. 

In contrast, the astronomers' reported redshift of 1.747 corresponds to emission lines of H$\gamma$, H$\beta$, the OIII doublet, H$\alpha$, NII, and the SII doublet, with H$\alpha$ (and the nearby, weaker NII line) as the prominent line feature. This reflects the common modeling assumption that the strongest observed line corresponds to H$\alpha$ or OIII, without fully accounting for the possibility of other elements when emission lines are weaker. The prior incorporates this belief, but does not enforce it deterministically as in previous fitting methods. Point estimates for the nuisance parameters $\alpha, \beta, \sigma^2,$ and $\delta$ are computed as described in Section~\ref{app:pointests} of the Supplementary Material, but are omitted here for brevity.

This spectrum illustrates that even when there is strong evidence for the presence of at least one emission line, substantial uncertainty may remain in the posterior distribution for redshift. The full analysis required approximately 15 seconds on a 20-core processor.

Next, we examine a posterior predictive plot to assess overall model fit. The results are shown in Figure~\ref{fig:post_pred_01}. The black dots represent the observed flux values. The light blue band corresponds to the pointwise 2.5\% and 97.5\% quantiles of 5000 draws from the posterior predictive distribution, and the dark blue line depicts the pointwise median of these draws. A full description of how posterior predictive samples are generated is provided in Section~\ref{app:ppc} of the Supplementary Material.

Visually, the model provides a good fit to the data. The posterior predictive bands capture most of the observed flux values without being overly wide. In particular, the model accurately reproduces the sharp peak associated with the most prominent emission line. On the right-hand side of the plot, the predictive bands widen substantially near the boundary. This widening reflects the measurement errors provided by JWST astronomers, rather than instability in the fitting procedure.

\subsubsection{Galaxy 00033}
We now analyze spectrum 00033 from the labeled JWST data. Figure~\ref{fig:spec_04} presents the corresponding posterior predictive plot. The log posterior odds in favor of $M_1$ are 120.2749, indicating extremely strong evidence for the presence of at least one emission line. In contrast to spectrum 00004, the marginal 99.865\% HPD set for redshift is the single interval $[1.867, 1.871]$, reflecting a unimodal marginal posterior, and the marginal MAP estimate is $1.869$. This is essentially identical to the astronomers' reported redshift of $1.86884$. At this redshift, the detected emission lines are the OII doublet, H$\gamma$, H$\beta$, the OIII doublet, H$\alpha$, NII, and the SII doublet. The corresponding posterior mean signal intensities (conditional on $\hat{\sigma}^2$ and $\hat{\zeta}$) are 0.0204, 0.0204, 0.0025, 0.0126, 0.0377, 0.1123, 0.0523, 0.0126, 0.0069, and 0.0093, respectively, with the most prominent feature corresponding to the OIII doublet. 

The posterior predictive fit is somewhat poorer than for spectrum 00004, as the predictive bands do not capture background and noise fluctuations as closely. Nevertheless, the presence of multiple emission lines tightly constrains the redshift. In this case, strong detection evidence coincides with precise redshift estimation.

\begin{figure}[t]
    \centering
    \includegraphics[width=\linewidth]{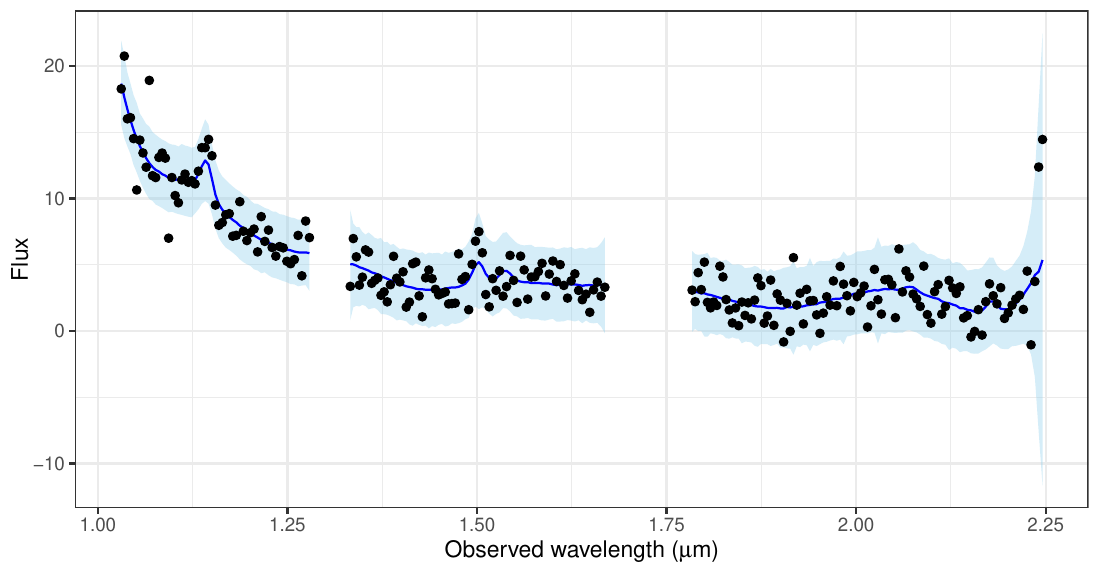}
    \caption{A posterior predictive plot assessing model fit for spectrum 02263. The black dots show the observed flux values per wavelength. Here, 5000 replicate datasets from the posterior predictive distribution are drawn, with the pointwise median per wavelength shown in dark blue, and the pointwise 2.5\% and 97.5\% pointwise quantiles as the lower and upper bounds of the light blue region, respectively.}
    \label{fig:spec_171}
\end{figure}

\begin{figure}[ht]
    \centering
    \includegraphics[width=\linewidth]{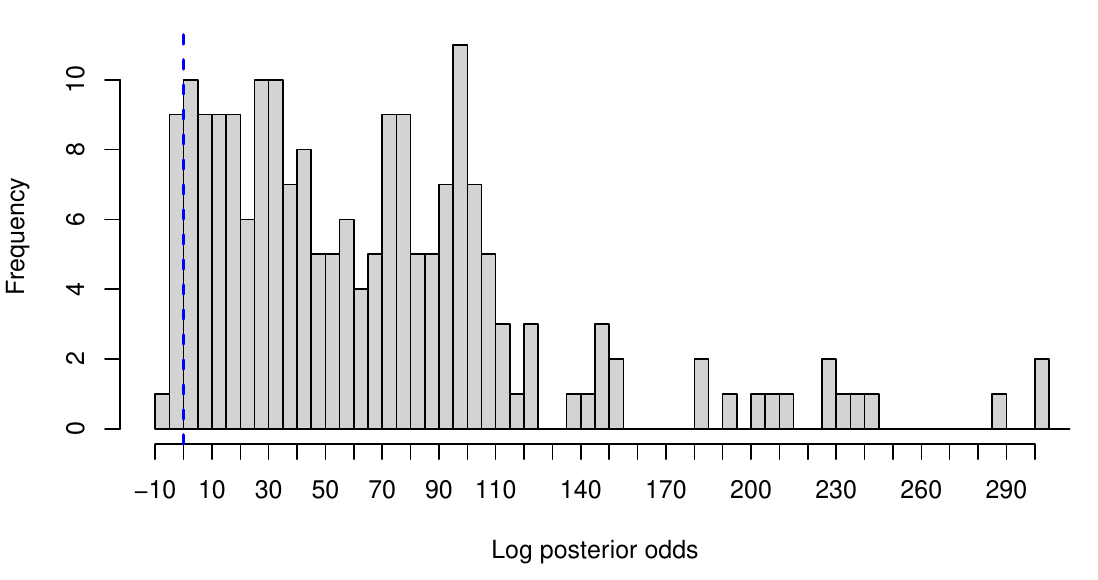}
    \caption{Histogram of log posterior odds for 213 labeled spectra. Some values are cutoff for being too large. The dashed blue line at zero represents the cutoff threshold for emission line detection.}
    \label{fig:log_PO_hist_labeled}
\end{figure}

\subsubsection{Galaxy 02263}
In contrast to the previously analyzed spectra, spectrum 02263 provides comparatively weak evidence for emission-line detection due to a prominent astrophysical background and faint emission lines. The posterior predictive plot is shown in Figure~\ref{fig:spec_171}. For this spectrum, the log posterior odds in favor of $M_1$ is 0.673, corresponding to a posterior probability of $M_1$ of 0.66. Thus, while at least one emission line is detected under the decision rule, the evidence is substantially weaker than in the previous two cases. The marginal MAP for redshift is $1.286$, which is again quite similar to the astronomers' provided redshift of $1.2854$. The corresponding 99.865\% marginal HPD set for redshift the union of 13 disjoint intervals (omitted for brevity), reflecting considerable uncertainty in the redshift estimate. 

\begin{figure}[ht]
    \centering
    \includegraphics[width=\linewidth]{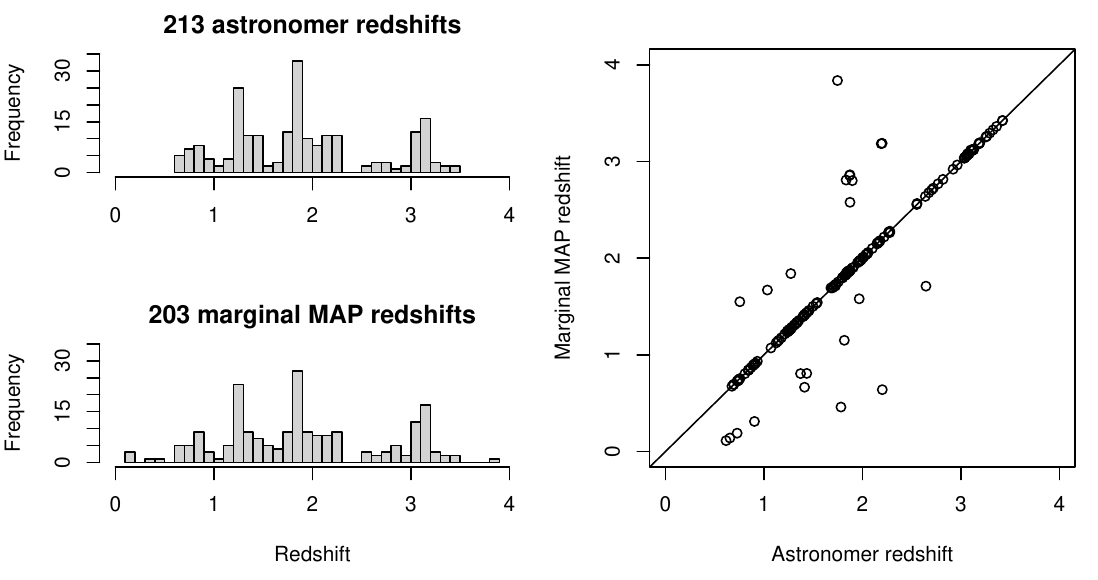}
    \caption{Left panel: Histograms of redshift estimates for labeled spectra from astronomers (top) and marginal MAP estimates (bottom). Counts differ because our selection procedure excludes ten spectra identified by astronomers based on the log posterior odds. Right panel: Scatterplot comparing astronomer and marginal MAP redshift estimates for the 203 spectra selected by the log posterior odds.}
    \label{fig:redshift_hists_labeled}
\end{figure}

In terms of model fit, the majority of observed flux values fall within the pointwise posterior predictive intervals, except for a few outliers at lower wavelengths. The overall mean structure is reasonably well captured, and multiple weak emission features are visible in the posterior predictive median curve. 

Taken together, these three examples demonstrate that redshift uncertainty is driven not only by the presence of emission lines but also by their number, relative strength, and interaction with the background structure. A single prominent feature may yield decisive evidence for detection while leaving redshift weakly identified. In contrast, multiple strong lines can tightly constrain redshift. When the signal strength is weak relative to the background and noise levels, both detection evidence and redshift precision deteriorate. These regimes recur throughout the JWST dataset and are reflected more broadly in the aggregate analysis that follows.

\subsection{Aggregate results for 213 labeled spectra}
\label{sec:agg_results_labeled}

We now consider the full set of 213 JWST spectra for which astronomers have identified emission lines and provided redshifts. Although these expert-provided annotations are not ground truth, they provide a natural point of comparison for our emission line detection decisions and redshift estimates.

Figure~\ref{fig:log_PO_hist_labeled} displays a histogram of log posterior odds across the 213 spectra. The blue vertical line marks zero, the decision threshold for emission line detection. The majority of spectra lie to the right of this line, with large positive log-posterior odds in favor of $M_1$.

Ten galaxy spectra are not selected as containing emission lines under the posterior odds decision rule. All 213 of these spectra were flagged by astronomers as likely to contain emission lines, so the high overall agreement is reassuring. Nevertheless, our procedure is more conservative than the expert labeling in this subset. This difference reflects, in part, the use of prior model probabilities for a generic set of galaxy spectra rather than for a pre-selected set likely to contain emission lines. When using the log Bayes factor alone (corresponding to equal prior model probabilities), only five spectra are not selected. 

\begin{figure}[ht]
    \centering
    \includegraphics[width=\linewidth]{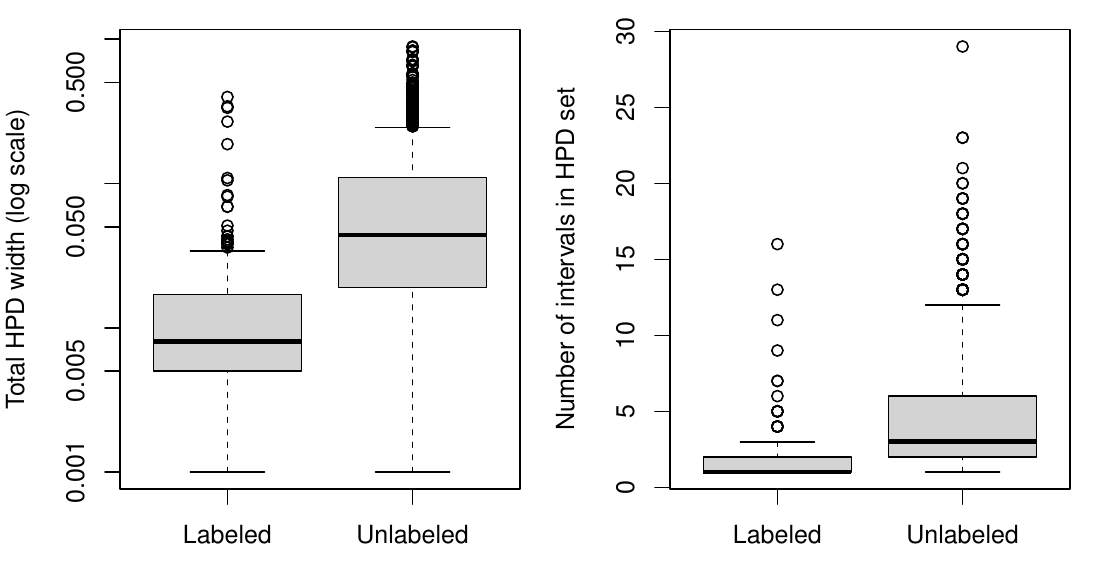}
    \caption{Left panel: Boxplots of the total width of 99.865\% HPD sets for spectra selected by our procedure as containing at least one emission line for the labeled (left) and unlabeled (right) JWST datasets. HPD sets are generally much narrower for the labeled data. Right panel: Boxplots of the number of disjoint intervals in each 99.865\% HPD set. Labeled spectra typically have 1–2 intervals, whereas unlabeled spectra more often have 3–6.}
    \label{fig:hpd_boxplots}
\end{figure}

\begin{figure}[ht]
    \centering
    \includegraphics[width=\linewidth]{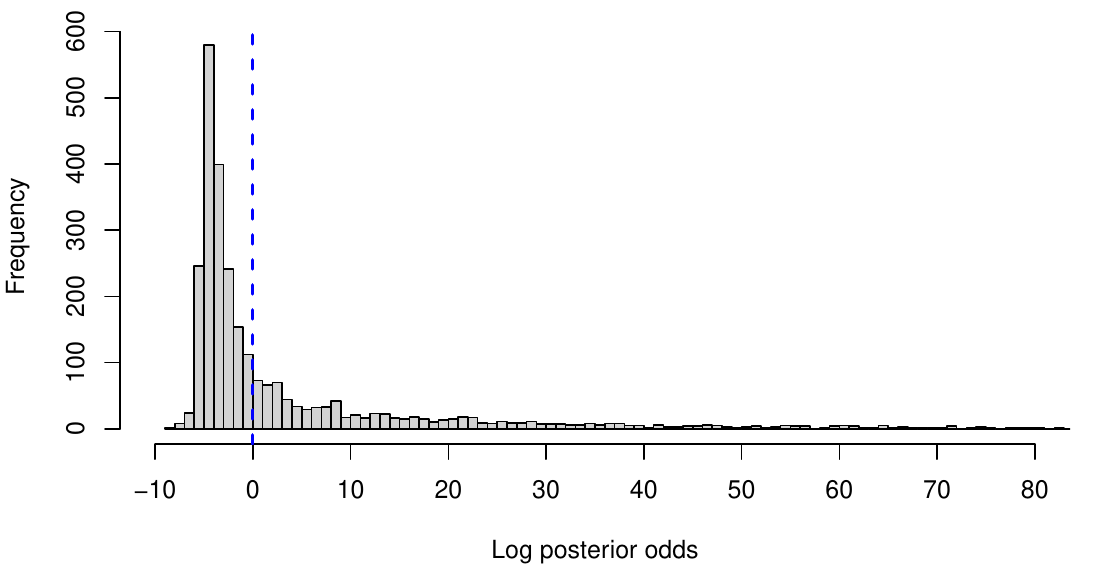}
    \caption{Histogram of log posterior odds for 2715 unlabeled spectra. The dashed blue line at zero represents the cutoff decision for line detection. Note the range on the x-axis is smaller compared to Figure~\ref{fig:log_PO_hist_labeled}.}
    \label{fig:log_PO_hist_unlabeled}
\end{figure}

Examination of the spectra not selected by our procedure indicates that they typically exhibit low signal-to-noise ratios or prominent background features relative to the emission line strengths (similar to spectrum 02263 above). In these cases, the redshift posterior is weakly constrained, so even a more liberal detection rule would not necessarily yield precise or informative redshift estimates. 

\subsection{Aggregate results for 2800 unlabeled spectra}
\label{sec:agg_results_unlabeled}
\begin{figure}[ht]
    \centering
    \includegraphics[width=\linewidth]{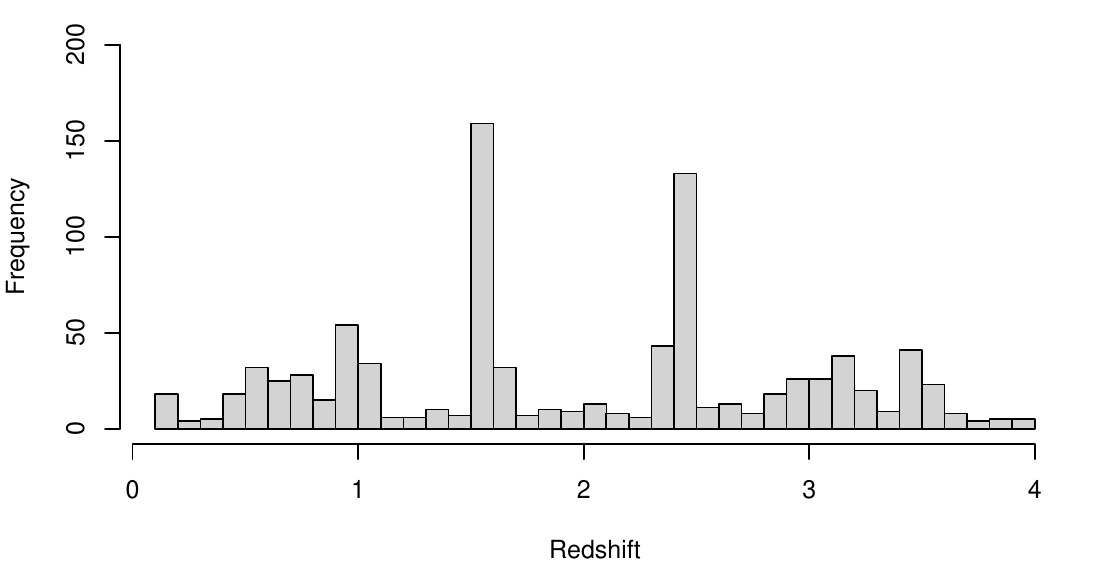}
    \caption{Histogram of redshift marginal MAP estimates for spectra in the unlabeled set that are selected according to posterior odds.}
    \label{fig:redshift_hist_unlabeled_selected}
\end{figure}

We estimate the redshift only in spectra for which emission lines are detected. The left panel of Figure~\ref{fig:redshift_hists_labeled} compares histograms of the marginal MAP redshift estimates from our procedure with the redshifts reported by astronomers. We emphasize again that neither set of values should be regarded as ground truth. Overall, the two distributions are similar. The most noticeable difference occurs in the low-redshift range: the astronomer-reported redshifts include no galaxies with redshifts between 0 and 0.5, whereas our procedure identifies eight galaxies in this range. The right panel of Figure~\ref{fig:redshift_hists_labeled} shows a corresponding scatterplot comparing redshift estimates for the 203 spectra selected by our procedure based on the log posterior odds. There is strong agreement between the two sets of estimates for most galaxies; however, 22 galaxies lie off the diagonal, indicating discrepancies in their fitted redshifts.

The left panel of Figure~\ref{fig:hpd_boxplots} summarizes the total widths of the marginal 99.865\% HPD sets for redshift for spectra selected as containing emission lines by our procedure. The distribution of total widths is highly skewed (note the log scale), with most labeled spectra having total widths between 0.005 and 0.02. For reference, recall that the redshift resolution used in the fitting procedure is 0.001. 

To better characterize the source of this uncertainty, the right panel of Figure~\ref{fig:hpd_boxplots} shows the number of disjoint intervals within each 99.865\% HPD set. For the labeled data, most spectra have HPD sets consisting of a single interval, or occasionally two, indicating that the posterior is typically unimodal or bimodal. A small number of spectra exhibit many disjoint intervals, corresponding to more weakly constrained redshift estimates. Thus, for the labeled sample, the width of the 99.865\% HPD sets generally reflects dispersion around a dominant redshift mode rather than the presence of several well-separated, competing redshift solutions. 

\begin{figure}[ht]
    \centering
    \includegraphics[width=\linewidth]{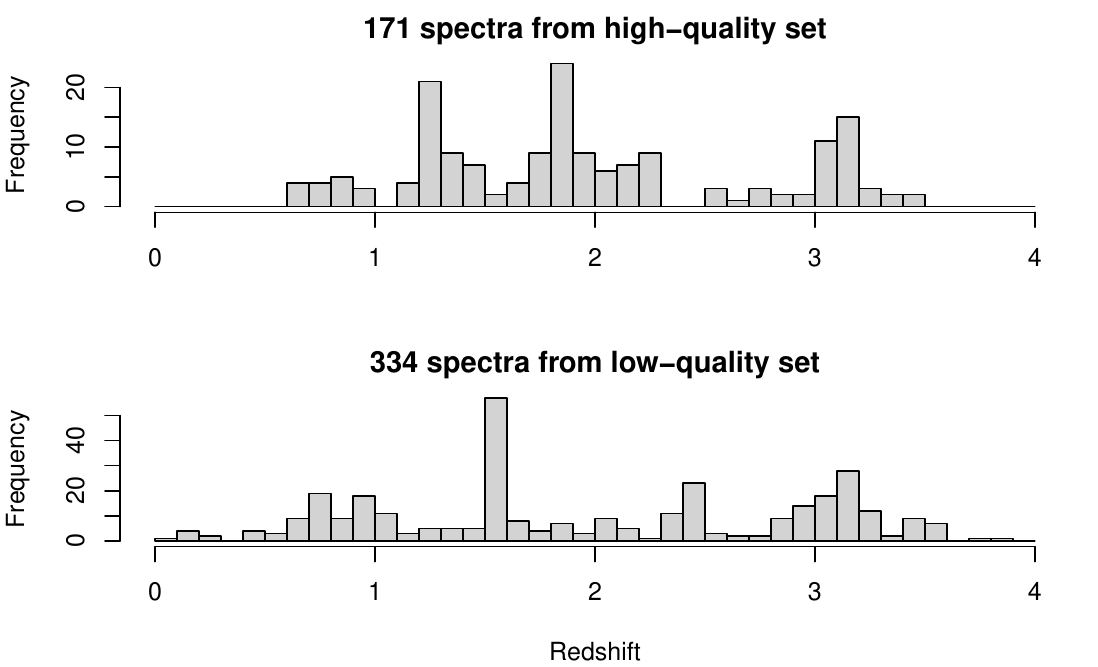}
    \caption{Comparison of marginal MAP redshift distributions for labeled (top) and unlabeled (bottom) spectra after restricting to spectra with strong evidence for emission line presence ($P(M_1 \mid y) > 0.99865$), and 99.865\% HPD sets consisting of one or two disjoint intervals. The dominant modes occur in different redshift regions across the two samples.}
    \label{fig:redshift_histogram_comparison}
\end{figure}

We now present aggregate results for the 2800 unlabeled JWST spectra. Due to the size of the dataset and the generally lower quality of these spectra, they have not been screened for emission line presence or assigned redshift estimates by astronomers. Within this set, some spectra exhibit data quality issues that render inference unreliable. These issues arise from various errors in the detection process, resulting in large numbers of extreme or missing values. The screening procedure used to select spectra of reasonable quality is described in Section~\ref{app:ppc} of the Supplementary Material and removes 85 spectra from the analysis. We therefore proceed using the remaining 2715 spectra.

Figure~\ref{fig:log_PO_hist_unlabeled} displays a histogram of log posterior odds for such spectra. The distribution is strongly right-skewed, with most values below zero (more extreme values are truncated in the figure). The lower quartile, median, and upper quartile are -4.3, -2.6, and 4.1, respectively, with a maximum value of 570. Under the posterior odds decision rule, emission lines are detected in 937 spectra, corresponding to approximately 35\% of the unlabeled sample.

Figure~\ref{fig:redshift_hist_unlabeled_selected} displays a histogram of marginal MAP redshift estimates for the 937 spectra identified as containing at least one emission line. The overall shape of the distribution is broadly similar to that observed for the labeled sample. However, the modes near redshift 1.6 and 2.5 are more pronounced here than in the high-quality set.

It is natural to ask whether spectra with particularly strong evidence for emission line presence, or those with more tightly constrained redshift estimates (e.g., HPD sets consisting of a small number of disjoint intervals), produce a distribution more comparable to that of the high-quality spectra. To examine this, we restrict attention to spectra with posterior probability of $M_1$ at least 0.99865 (the $3\sigma$ probability, corresponding to a log posterior odds of at least 6.6), and whose 99.865\% HPD set for redshift consists of either one or two disjoint intervals. We allow up to two intervals in order to retain a reasonable number of observations from the high-quality sample, whose size is limited. 

The filtered histograms for both the labeled and unlabeled spectra are shown in Figure~\ref{fig:redshift_histogram_comparison}.
The resulting distributions differ substantially. In particular, the dominant modes occur in distinct redshift regions across the two samples. Notably, the boundary redshift values observed in the high-quality set disappear entirely when restricting attention to spectra with strong evidence for emission line presence.

We next examine posterior uncertainty in redshift estimation. Figure~\ref{fig:hpd_boxplots} summarizes posterior uncertainty for redshift in the unlabeled spectra (and labeled spectra as discussed in the previous section). The total widths of the 99.865\% HPD sets are generally larger than those observed in the labeled sample, and the number of disjoint intervals is correspondingly higher, indicating greater posterior uncertainty in redshift estimation. Nevertheless, 159 unlabeled spectra yield HPD sets consisting of a single interval, suggesting well-constrained redshift estimates for a subset of galaxies that is nearly as large as the expert-labeled dataset. 

Moreover, the posterior odds provide a natural ranking of spectra by the strength of evidence for emission-line detection. This ranking enables targeted follow-up analysis: spectra with large posterior odds can be prioritized for more sophisticated modeling where computational cost is less restrictive, or post-hoc expert investigations. Processing all 2800 spectra required under 12 hours using modest computational resources (a 20-core processor and 32 GB of RAM). Because the fitting procedure admits parallel computation within each spectrum, the approach scales naturally to much larger datasets.

\section{Final Remarks}
\label{sec:final_remarks}
We have presented a modeling framework and analysis of approximately 3,000 JWST spectra, providing a structured procedure for simultaneously detecting multiple emission lines and quantifying the uncertainty in redshift estimation. The approach explicitly accounts for systematic uncertainties arising from background modeling, excess variance beyond the reported measurement error, and emission line widths. Our analysis suggests that redshift uncertainty may be substantially larger than is typically reported, even for some high-quality spectra with expert labels.  

The results exhibit sensitivity to the choice of some prior hyperparameters. In particular, the number of background basis functions $J$ should be chosen sufficiently large to allow the null model adequate flexibility (see Section~\ref{app:bg} of the Supplementary Material). The value $J = 15$ used in Section~\ref{sec:results} appears appropriate for this purpose. Additionally, the log posterior odds and the distribution of marginal MAP redshift estimates are sensitive to the prior specification for the signal intensities (Section~\ref{app:sig}). Priors that incorporate astrophysical knowledge of relative emission line strengths lead to more stable redshift inference, whereas diffuse intensity priors can substantially alter the marginal MAP redshift distribution. This sensitivity is amplified by using the marginal MAP as a point estimate, which depends on the relative heights of modes in the marginal posterior. In contrast, the results show relatively little sensitivity to the hyperparameters governing the redshift prior (Section~\ref{app:redshift}).

We recommend reporting HPD sets alongside marginal MAP estimates in downstream analyses whenever possible. By capturing posterior multimodality, HPD sets are generally less sensitive to prior hyperparameter choices than the marginal MAP estimate alone.

To achieve computational feasibility at the scale required by modern spectroscopic surveys such as JWST, Euclid, and Roman, several modeling compromises were necessary. These include distributional assumptions in the likelihood (e.g., normally distributed errors), the use of truncated multivariate normal priors for signal intensities, and a plug-in estimate for $\sigma^2$ in the covariance structure. These simplifications enable large-scale analysis, but limit full Bayesian uncertainty propagation. 
Nevertheless, the proposed methodology appears to perform well in our data analysis in Section~\ref{sec:results} and in the numerical experiments (Sections~\ref{app:model_check} and~\ref{app:sens} of the Supplementary Material).

For analyses focused on specific galaxies, or refined subsets where highly precise redshift estimates are required, a fully Bayesian treatment would be desirable. Even for a single spectrum, however, the posterior distribution can be highly multimodal. As illustrated in our analysis, a prominent emission feature may induce multiple competing redshift modes. Consequently, advanced MCMC methods such as parallel tempering and related tempered approaches \citep{geyerAnnealingMarkovChain1995, syedNonReversibleParallelTempering2022} would likely be required. Developing such algorithms for models with varying numbers of active emission lines and associated constraints remains an important methodological challenge. In addition, complementary frequentist procedures for emission line detection and redshift estimation would provide useful alternative inferential tools for large spectroscopic surveys.

\begin{funding}
AK, SA, and GJ were partially supported by NSF grant DMS-2152746. SA was partially supported by the Warwick Mid-Career Award, CLA, University of Minnesota. AK was partially supported by the NSF Grant NRT-1922512. CS knowledges the support of NASA ROSES Grant 12-EUCLID11-0004.
\end{funding}

\begin{supplement}
\stitle{Supplement to ``A scalable Bayesian framework for Galaxy emission line detection and redshift estimation''}
\sdescription{Additional derivations, descriptions of optimization and sampling schemes, model fit assessments, and a sensitivity analysis.}
\end{supplement}
\begin{supplement}
\stitle{Data and R Code}
\sdescription{A subset of the full JWST spectral data, corresponding astronomer redshift annotations, and R code required to reproduce all results and figures reported in the paper. Materials also available at \url{https://github.com/alexkuhn0115/Bayesian-emission-line-detection-public}.}
\end{supplement}


\bibliographystyle{imsart-nameyear} 
\bibliography{bibliography}


\makeatletter
\protected@write\@auxout{}{%
  \string\newlabel{LastEquation}{{\number\value{equation}}{}{}{}{}}}
\makeatother

\ifarXiv
\renewcommand{\appendixname}{Supplement}
\renewcommand{\thesection}{S.\arabic{section}}
\renewcommand{\theequation}{S.\arabic{equation}}
\renewcommand{\thefigure}{S.\arabic{figure}}
\renewcommand{\thetable}{S.\arabic{table}}

\setcounter{section}{0}
\setcounter{equation}{0}
\setcounter{figure}{0}
\setcounter{table}{0}

\clearpage
\begin{center}
{\large SUPPLEMENT TO ``A SCALABLE BAYESIAN FRAMEWORK FOR EMISSION LINE DETECTION AND REDSHIFT ESTIMATION''}
\end{center}
\input{supp_content}
\fi

\end{document}

%% file: supp_content.tex
\section{Likelihood derivation}
\label{app:orthog}
Recall that we impose orthogonality conditions on the linear model in equation~\eqref{eq:model} in the main text. In particular, we require
\begin{align*}
       1_N'[S(\sigma^2)]^{-1}X &= 0,\\
       1_N'[S(\sigma^2)]^{-1}W_{\text{sub}}(\zeta, \delta) &= 0,\\
       X'[S(\sigma^2)]^{-1}W_{\text{sub}}(\zeta, \delta) &= 0,
\end{align*}
where the zero-vector on the right-hand side of each equation above is of appropriate dimension. To do so, consider the projection matrix onto the span of $1_N$ as 
\begin{equation*}
    P_{1_N} = 1_N(1_N'[S(\sigma^2)]^{-1}1_N)^{-1}1_N'[S(\sigma^2)]^{-1},
\end{equation*}
and let $M_{1_N} = I_N - P_{1_N}$ denote the corresponding orthogonal complement projection matrix. Projecting $X$ using this operator gives $X_{\perp} = M_{1_N}X,$ which satisfies \begin{equation*}
    1_N'[S(\sigma^2)]^{-1}X_{\perp} = 0.
\end{equation*}
Next, define the projection matrix onto the span of $X_{\perp}$ as 
\begin{equation*}
    P_{X_{\perp}} = X_{\perp}(X_{\perp}'[S(\sigma^2)]^{-1} X_{\perp})^{-1}X_{\perp}'[S(\sigma^2)]^{-1},
\end{equation*}
and the corresponding orthogonal complement as $M_{X_{\perp}} = I_N - P_{X_{\perp}}$. Then, for each $(\zeta, \delta)$, the orthogonalized matrix $W_{\text{sub}, \perp}(\zeta, \delta) = M_{X_{\perp}}M_{1_N}W_{\text{sub}}(\zeta, \delta)$ satisfies the orthogonality conditions 
\begin{align*}
1_N'[S(\sigma^2)]^{-1}W_{\text{sub},\perp}(\zeta,\delta) &= 0,\\
X_{\perp}'[S(\sigma^2)]^{-1}W_{\text{sub},\perp}(\zeta,\delta) &= 0.
\end{align*}
For simplicity of notation, we write $X$ and $W_{\text{sub}}$ rather than $X_{\perp}$ and $W_{\text{sub}, \perp}$, bearing in mind that this orthogonalization has been performed. 

With the orthogonality conditions above, the generalized least-squares estimators of $\alpha, \beta$ and $\eta_{\text{sub}}(\zeta)$ for fixed $(\zeta, \delta)$ are
\begin{align} 
    \hat{\alpha} &= (1'_N[S(\sigma^2)]^{-1}1_N)^{-1}1_N'[S(\sigma^2)]^{-1}y, \label{eq:lsalpha}\\
    \hat{\beta} &= (X'[S(\sigma^2)]^{-1}X)^{-1}X'[S(\sigma^2)]^{-1}(y - \hat{\alpha}1_N), \label{eq:lsbeta}\\
    \hat{\eta}_{\text{sub}}(\zeta) &= [W_{\text{sub}}(\zeta, \delta)'[S(\sigma^2)]^{-1}W_{\text{sub}}(\zeta, \delta)]^{-1}W_{\text{sub}}(\zeta, \delta)'[S(\sigma^2)]^{-1}(y - \hat{\alpha}1_N - X\hat{\beta}). \label{eq:lseta}
\end{align} 
Let $r = y - \hat{\alpha}1_N - X\hat{\beta}$. A routine calculation \citep[see, e.g., Ch. 3][]{marinBayesianEssentials2014} yields
\begin{equation}
\label{eq:likelihood}
\begin{aligned}
    p(y \mid \alpha, \beta, \sigma^2, \eta_{\text{sub}}(\zeta), \zeta, \delta)
    &\propto \exp\Bigg\{-\frac{1}{2}(\eta_{\text{sub}}(\zeta) - \hat{\eta}_{\text{sub}}(\zeta))'\\
    &\qquad \qquad \times W_{\text{sub}}(\zeta,\delta)'[S(\sigma^2)]^{-1}W_{\text{sub}}(\zeta,\delta)(\eta_{\text{sub}}(\zeta) - \hat{\eta}_{\text{sub}}(\zeta))\Bigg\} \\
    &\quad\times \exp\left\{-\frac{1}{2}(\alpha - \hat{\alpha})1_N'[S(\sigma^2)]^{-1}1_N(\alpha - \hat{\alpha})\right\} \\
    &\quad\times \exp\left\{-\frac{1}{2}(\beta - \hat{\beta})'X'[S(\sigma^2)]^{-1}X(\beta - \hat{\beta})\right\} \\
    &\quad\times \exp\Bigg\{-\frac{1}{2}[r - W_{\text{sub}}(\zeta,\delta)\hat{\eta}_{\text{sub}}(\zeta)]'\\
    &\qquad \qquad \quad \times [S(\sigma^2)]^{-1}[r - W_{\text{sub}}(\zeta,\delta)\hat{\eta}_{\text{sub}}(\zeta)]\Bigg\}\\
    &\quad\times (\sigma^2)^{-N/2} 
\end{aligned}
\end{equation}
where the cross-terms are zero by the orthogonal construction and properties of least-squares estimates (and the prior on $\delta$ is a constant by uniformity). From the full model likelihood in equation~\eqref{eq:likelihood}, we can obtain the likelihood for the null model by setting $\eta_{\text{sub}}(\zeta) = \hat{\eta}_{\text{sub}}(\zeta) = 0$. 

\section{Marginal density under alternative model}\label{app:deriv}

This section provides details for computing the density $p(y \mid \hat{\sigma}^2, \zeta, \delta)$ described in Section~\ref{sec:integration}. To do so, we consider the encompassing model defined in equation~\eqref{eq:encomp}. 

\subsection{Encompassing model derivation}
\label{app:marg_deriv}
In this setting, we have the (trivial but key) relationship
\begin{equation*}
    p(\eta_{\text{sub}}(\zeta) \mid \zeta) = \frac{p(\eta^*_{\text{sub}}(\zeta) \mid \zeta)1_{\{\eta^*_{\text{sub}}(\zeta) \in \mathcal{A}(\zeta)\}}}{\Pr(\eta^*_{\text{sub}}(\zeta) \in \mathcal{A}(\zeta) \mid \zeta)},
\end{equation*}
with $\mathcal{A}(\zeta)$ defined as in \eqref{eq:cons} and with $\eta^*_{\text{sub}}(\zeta)$ satisfying \eqref{eq:encomp}. Using the encompassing model, equation~\eqref{eq:pstar_density} gives an expression for $p^*(y \mid \hat{\sigma}^2, \zeta, \delta)$ after integrating out $\alpha, \beta,$ and $\eta_{\text{sub}}^*(\zeta)$ directly due to conjugacy. Using this expression, we can integrate out $\alpha, \beta,$ and $\eta_{\text{sub}}(\zeta)$ to obtain $p(y \mid \hat{\sigma}^2, \zeta, \delta)$. First, $\alpha$ and $\beta$ can be integrated out directly by conjugacy, so $p(y \mid \hat{\sigma}^2, \eta_{\text{sub}}(\zeta), \zeta, \delta)$ is just the density of a multivariate normal distribution. Next, $\eta_{\text{sub}}(\zeta)$ can be integrated out as follows:
\begin{align*}
    p(y \mid \hat{\sigma}^2, \zeta, \delta) &= \int p(y \mid \hat{\sigma}^2, \eta_{\text{sub}}(\zeta),\zeta, \delta)p(\eta_{\text{sub}}(\zeta) \mid \zeta)d\eta_{\text{sub}}(\zeta)\\
    &= \int p(y \mid \hat{\sigma}^2, \eta^*_{\text{sub}}(\zeta), \zeta, \delta) \frac{p(\eta^*_{\text{sub}}(\zeta) \mid \zeta)1_{\{\eta^*_{\text{sub}}(\zeta) \in \mathcal{A}(\zeta)\}}}{\Pr(\eta^*_{\text{sub}}(\zeta) \in \mathcal{A}(\zeta) \mid \zeta)} d\eta^*_{\text{sub}}(\zeta)\\
    &= \frac{p^*(y \mid \hat{\sigma}^2, \zeta, \delta)}{\Pr(\eta^*_{\text{sub}}(\zeta) \in \mathcal{A}(\zeta) \mid \zeta)} \int p(\eta^*_{\text{sub}}(\zeta) \mid \hat{\sigma}^2, \zeta, \delta, y)1_{\{\eta^*_{\text{sub}}(\zeta) \in \mathcal{A}(\zeta)\}}d\eta^*_{\text{sub}}(\zeta)\\
    &= p^*(y \mid \hat{\sigma}^2, \zeta, \delta)\frac{\Pr(\eta^*_{\text{sub}}(\zeta) \in \mathcal{A}(\zeta) \mid \hat{\sigma}^2, \zeta, \delta, y)}{\Pr(\eta^*_{\text{sub}}(\zeta) \in \mathcal{A}(\zeta) \mid \zeta)},
\end{align*}
where 
\begin{equation*}
    p^*(y \mid \hat{\sigma}^2, \zeta, \delta) = \int p(y \mid \hat{\sigma}^2, \eta^*_{\text{sub}}(\zeta), \zeta, \delta) p(\eta^*_{\text{sub}}(\zeta) \mid \zeta)\,d\eta^*_{\text{sub}}(\zeta).
\end{equation*}
From the derivation in Appendix~\ref{app:eta_posterior}, it follows that $\eta^*_{\text{sub}}(\zeta) \mid \sigma^2, \zeta, \delta, y$ follows a multivariate normal distribution with known mean and covariance matrix. Next, we demonstrate how to quickly compute 
\begin{equation}
\label{eq:probs}
    \Pr(\eta^*_{\text{sub}}(\zeta) \in \mathcal{A}(\zeta) \mid\sigma^2, \zeta, y) 
\end{equation}
and $\Pr(\eta^*_{\text{sub}}(\zeta) \in \mathcal{A}(\zeta) \mid \zeta)$ in R. 

\subsection{Linear inequality constraints}
The constraints in the set $\mathcal{A}(\zeta)$ can always be written as $A(\zeta) \eta^*_{\text{sub}}(\zeta) \geq 0$ for an appropriate matrix, $A(\zeta)$. To construct the latter, note that the nonnegativity and ratio constraints in equation~\eqref{eq:cons} are partly redundant. We therefore retain only a minimal set of inequalities defining the same feasible region for fixed $\zeta$. Whenever two or more signal intensities (the $\eta_k$'s) are related by ratio constraints, keep only the nonnegativity constraint for the first index in that ordered set (i.e. the smallest of $k = 3, 4, 7$ in $\mathcal{K}(\zeta)$), and include only the pairwise ratio constraints comparing each subsequent element to the previous one. For example, when $\mathcal{K}(\zeta) = \{3, 4, 5, 6, 7\}$, the active constraints can be written as
\begin{equation*}
    \eta_4(\zeta) - 2.174\eta_3(\zeta) \geq 0 ,\quad \eta_7(\zeta) - 2.86\eta_4(\zeta) \geq 0, \quad \eta_k(\zeta) \geq 0, \; k \in \mathcal{K}(\zeta).
\end{equation*}
The reduced, nonredundant system is therefore
\begin{equation*}
\begin{aligned}
    &\eta_4(\zeta) - 2.174\,\eta_3(\zeta) \geq 0, \quad
    \eta_7(\zeta) - 2.86\,\eta_4(\zeta) \geq 0, \\
    &\eta_3(\zeta) \geq 0, \quad 
    \eta_5(\zeta) \geq 0, \quad 
    \eta_6(\zeta) \geq 0,
\end{aligned}
\end{equation*}
which we can express compactly in matrix form as $A(\zeta)\eta_{\text{sub}}(\zeta) \geq 0$, where
\begin{equation*}
    A(\zeta) =
    \begin{bmatrix}
        -2.174 & 1 & 0 & 0 & 0 \\[4pt]
        0 & -2.86 & 0 & 0 & 1 \\[4pt]
        1 & 0 & 0 & 0 & 0 \\[4pt]
        0 & 0 & 1 & 0 & 0 \\[4pt]
        0 & 0 & 0 & 1 & 0
    \end{bmatrix},
    \qquad
    \eta_{\text{sub}}(\zeta) =
    \begin{bmatrix}
        \eta_3 \\[4pt]
        \eta_4 \\[4pt]
        \eta_5 \\[4pt]
        \eta_6 \\[4pt]
        \eta_7
    \end{bmatrix}.
\end{equation*}
Since $A(\zeta)\eta^*_{\text{sub}}(\zeta) \mid \hat{\sigma}^2, \zeta, y$ follows a multivariate normal distribution with mean $A(\zeta)m_{\hat{\sigma}^2, \hat{\eta}_{\text{sub}}(\zeta), \delta}$ and covariance matrix $A(\zeta)V_{\hat{\sigma}^2, \hat{\eta}_{\text{sub}}(\zeta), \delta}A(\zeta)'$, where $m_{\hat{\sigma}^2, \hat{\eta}_{\text{sub}}(\zeta), \delta}$ and $V_{\hat{\sigma}^2, \hat{\eta}_{\text{sub}}(\zeta), \delta}$ are the centrality and spread parameters for the truncated multivariate normal posterior of $p(\eta_{\text{sub}}(\zeta) \mid \hat{\sigma}^2, \zeta, \delta, y)$ defined in equations~\eqref{eq:postmean} and~\eqref{eq:postvar}, respectively. Importantly, these quantities are available in closed form. The probability in \eqref{eq:probs} corresponds to the nonnegative orthant probability for $A(\zeta)\eta^*_{\text{sub}}(\zeta)$, which can be evaluated directly using \texttt{pmvnorm()} in R. The same can be done using the prior mean and covariance instead.

Likewise, from the prior in equation~\eqref{eq:encomp}, we know that $A(\zeta)\eta^*_{\text{sub}}(\zeta) \mid \zeta$ follows a multivariate normal distribution as well with mean $A(\zeta)a_2(\zeta)$ and covariance matrix $A(\zeta)B_2(\zeta)A(\zeta)'$, where $a_2(\zeta)$ and $B_2(\zeta)$ are as defined in equation~\eqref{eq:hyperparams}.

\section{Posterior expectations}
\label{app:pointests}
In this section we derive the posterior mean for $\alpha, \beta, \delta,$ and $\eta_{\text{sub}}(\zeta)$, conditional on $\sigma^2 = \hat{\sigma}^2$ and $\zeta = \hat{\zeta} = \arg\max_{r}p(\zeta_r \mid y, \hat{\sigma}^2, M_1).$ 

\subsection{Background components}

From the likelihood derivation in Appendix~\ref{app:deriv}, we write
\begin{align*}
    p(\alpha \mid \hat{\sigma}^2, \hat{\zeta}, y) 
    &\propto \exp\left\{-\frac{1}{2}(\alpha - \hat{\alpha})1_N'[S(\hat{\sigma}^2)]^{-1}1_N(\alpha - \hat{\alpha})\right\} p(\alpha) \\
    &\propto \exp\left\{-\frac{1}{2}(\alpha - \hat{\alpha})1_N'[S(\hat{\sigma}^2)]^{-1}1_N(\alpha - \hat{\alpha})\right\} \exp\left\{-\frac{1}{2 b_0^2}(\alpha - a_0)^2\right\}\\
    &\propto \exp\left\{-\frac{1}{2 s^2_{\alpha}} (\alpha - m_{\alpha})^2\right\},
\end{align*}
by completing the square, where $\hat{\alpha}$ is defined in equation~\eqref{eq:lsalpha}, and
\begin{equation*}
    s_{\alpha}^2 = \left(1_N'[S(\hat{\sigma}^2)]^{-1}1_N + \frac{1}{b_0^2}\right)^{-1}, \quad    m_{\alpha} = s_{\alpha}^2\left(\hat{\alpha}1_N'[S(\hat{\sigma}^2)]^{-1}1_N + \frac{a_0}{b_0^2}\right).
\end{equation*}
In particular, $\alpha \mid \hat{\sigma}^2, \hat{\zeta}, y  \equiv \alpha \mid \hat{\sigma}^2, y\sim N(m_{\alpha}, s_{\alpha})$. The derivation for $\beta \mid \sigma^2, \hat{\zeta}, y$ is similar. We have
\begin{align*}
    p(\beta \mid \hat{\sigma}^2, \hat{\zeta}, y) 
    &\propto \exp\left\{-\frac{1}{2}(\beta - \hat{\beta})'X'[S(\hat{\sigma}^2)]^{-1}X(\beta - \hat{\beta})\right\}\exp\left\{-\frac{1}{2 b_1^2}\beta'\beta\right\}\\
    &\propto \exp\left\{(\beta - m_{\beta})' S_{\beta}^{-1}(\beta - m_{\beta})\right\},
\end{align*}
where $\hat{\beta}$ is defined in equation~\eqref{eq:lsbeta}, and 
\begin{equation*}
    S_{\beta} = \left(X'[S(\hat{\sigma}^2)]^{-1}X + \frac{1}{b_1^2}I_J\right)^{-1}, \quad     m_{\beta} = S_{\beta}^{-1}\left(X'[S(\hat{\sigma}^2)]^{-1}X\hat{\beta}\right),
\end{equation*}
where $I_J$ is the $J\times J$ identity matrix. That is, $\beta \mid \hat{\sigma}^2, \hat{\zeta}, \delta, y \equiv \beta \mid \hat{\sigma}^2, y \sim N_J(m_\beta, S_\beta)$.

\subsection{Emission line profile width}
The posterior expectation for $\delta$ is obtained as follows. For fixed $\hat{\sigma}^2, \hat{\zeta}$, the posterior probability of each $\delta_\ell$ is
\begin{equation}
\label{eq:post_delta}
    p(\delta_\ell \mid \hat{\sigma}^2, \hat{\zeta}, y) = \frac{p(y \mid \hat{\sigma}^2, \hat{\zeta}, \delta_\ell)p(\delta_\ell)\Delta_{\delta}}{\sum_{\ell = 1}^D p(y \mid \hat{\sigma}^2, \hat{\zeta},\delta_\ell)p(\delta_\ell)\Delta_{\delta}},
\end{equation}
where each $p(y \mid \hat{\sigma}^2, \hat{\zeta}, \delta_\ell)$ term is calculated according to equation~\eqref{eq:ev1}. Therefore, the posterior conditional mean of $\delta$ is
\begin{equation*}
    E(\delta \mid \hat{\sigma}^2, \hat{\zeta}, y) = \sum_{\ell=1}^D \delta_\ell p(\delta_\ell \mid \hat{\sigma}^2, \hat{\zeta}, y).
\end{equation*}

\subsection{Signal intensities}
\label{app:eta_posterior}
Next, we compute the conditional posterior mean, $E(\eta_{\text{sub}}(\hat{\zeta}) \mid \hat{\sigma}^2, \hat{\zeta}, y)$. To start, we derive the full conditional for $\eta_{\text{sub}}(\zeta)$ when the prior on $\eta_{\text{sub}} \mid \zeta$ is assumed to be the truncated multivariate normal defined in equation~\eqref{eq:tnorm}. Using the likelihood in equation~\eqref{eq:likelihood}, expressed in terms of the weighted least-squares estimate $\hat{\eta}_{\text{sub}}(\zeta)$, the conditional posterior density of $\eta_{\text{sub}}(\zeta)$ given $\hat{\sigma}^2$, $\zeta$, and $\delta$ can be written as
\begin{equation*}
\begin{aligned}
    p(\eta_{\text{sub}}(\zeta) \mid \hat{\sigma}^2, \zeta, \delta, y) 
    &\propto 
    \exp\Bigg\{-\frac{1}{2} (\eta_{\text{sub}}(\zeta) - \hat{\eta}_{\text{sub}}(\zeta))'\\
    &\qquad \qquad \times W_{\text{sub}}(\zeta, \delta)' S(\hat{\sigma}^2)^{-1}W_{\text{sub}}(\zeta, \delta) (\eta_{\text{sub}}(\zeta) - \hat{\eta}_{\text{sub}}(\zeta)) \Bigg\} \\
    &\quad \times 
    \exp\left\{-\frac{1}{2} (\eta_{\text{sub}}(\zeta) - a_2(\zeta))' B_2(\zeta)^{-1} (\eta_{\text{sub}}(\zeta) - a_2(\zeta)) \right\}\\
    &\quad \times 1_{\{\eta_{\text{sub}}(\zeta) \in \mathcal{A}(\zeta)\}} \\
    &\propto \exp\Bigg\{-\frac{1}{2} (\eta_{\text{sub}}(\zeta) - m_{\hat{\sigma}^2, \hat{\eta}_{\text{sub}}(\zeta), \delta})' 
    V_{\hat{\sigma}^2, \hat{\eta}_{\text{sub}}(\zeta), \delta}^{-1}\\
    &\qquad \qquad \times (\eta_{\text{sub}}(\zeta) - m_{\hat{\sigma}^2, \hat{\eta}_{\text{sub}}(\zeta), \delta}) \Bigg\}1_{\{\eta_{\text{sub}}(\zeta) \in \mathcal{A}(\zeta)\}},
\end{aligned}
\end{equation*}
where 
\begin{align}
\label{eq:postmean} 
    m_{\hat{\sigma}^2, \hat{\eta}_{\text{sub}}(\zeta), \delta} &:= V_{\hat{\sigma}^2, \hat{\eta}_{\text{sub}}(\zeta), \delta} \left(W_{\text{sub}}(\zeta, \delta)'[S(\hat{\sigma}^2)]^{-1}W_{\text{sub}}(\zeta, \delta)\hat{\eta}_{\text{sub}}(\zeta) + B_2^{-1}(\zeta)a_2(\zeta)\right),
\end{align}
and
\begin{equation}
\label{eq:postvar}
    V_{\hat{\sigma}^2, \hat{\eta}_{\text{sub}}(\zeta), \delta} := \left(W_{\text{sub}}(\zeta, \delta)'[S(\hat{\sigma}^2)]^{-1}W_{\text{sub}}(\zeta, \delta) + B_2^{-1}(\zeta)\right)^{-1}.
\end{equation}
Due to the restriction that $\eta_{\text{sub}}(\zeta) \in \mathcal{A}(\zeta)$, the posterior mean of $\eta_{\text{sub}}(\hat{\zeta}) \mid \hat{\sigma}^2, \hat{\zeta}, y$ is not simply $m_{\hat{\sigma}^2, \hat{\eta}_{\text{sub}}(\hat{\zeta}), \delta}$ (likewise for $V_{\hat{\sigma}^2, \hat{\eta}_{\text{sub}}(\hat{\zeta}), \delta}$ and the posterior variance). Instead, we consider the linear inequality constraints in equation~\eqref{eq:cons}, and need to compute
\begin{equation}
\label{eq:eta_post_mean}
    E[\eta_{\text{sub}}(\hat{\zeta}) \mid \hat{\sigma}^2, \hat{\zeta}, \delta, y] = E[\eta^*_{\text{sub}}(\hat{\zeta}) \mid A(\hat{\zeta})\eta^*_{\text{sub}}(\hat{\zeta}) \geq 0, \hat{\sigma}^2, \hat{\zeta}, \delta,y],
\end{equation}
where $\eta^*_{\text{sub}}(\hat{\zeta})$ denotes the corresponding non-truncated multivariate normal random variable in equation~\eqref{eq:encomp}. To simplify notation, we will suppress the conditioning on $\hat{\sigma}^2, \hat{\zeta},$ and $y$, since they are fixed in what follows, and write $T(\hat{\zeta}) = A(\hat{\zeta})\eta^*_{\text{sub}}(\hat{\zeta})$, $\hat{m}_{\delta} = m_{\hat{\sigma}^2, \hat{\eta}_{\text{sub}}(\hat{\zeta}), \delta}$, and $\hat{V}_{\delta} = V_{\hat{\sigma}^2, \hat{\eta}_{\text{sub}}(\hat{\zeta}), \delta}$. The right-hand side of~\eqref{eq:eta_post_mean} can then be expressed as $E[\eta^*_{\text{sub}}(\hat{\zeta}) \mid T(\hat{\zeta}) \geq 0, \delta]$. To evaluate this expectation, note that $\eta^*_{\text{sub}}(\hat{\zeta})$ and $T(\hat{\zeta})$ are jointly multivariate normal (conditional on $\delta$) with mean vector $(\hat{m}_{\delta}, A(\hat{\zeta})\hat{m}_{\delta})'$
and covariance matrix
\begin{equation*}
    \begin{bmatrix}
        \hat{V}_{\delta} & \hat{V}_{\delta}A(\hat{\zeta})' \\
        A(\hat{\zeta})\hat{V}_{\delta} & A(\hat{\zeta})\hat{V}_{\delta}A(\hat{\zeta})'
    \end{bmatrix}.
\end{equation*}
Let $u \in \mathbb{R}$ denote an arbitary fixed value of $T(\hat{\zeta})$. For any such $u$, it can be shown \citep[e.g., Theorem 4.4d,][]{rencherLinearModelsStatistics2008a} that the conditional mean  is
\begin{equation*}
    E[\eta^*_{\text{sub}}(\hat{\zeta}) \mid T(\hat{\zeta}) = u, \delta] = \hat{m}_{\delta} + \hat{V}_{\delta}A(\hat{\zeta})'[A(\hat{\zeta})\hat{V}_{\delta}A(\hat{\zeta})']^{-1}[u - A(\hat{\zeta})\hat{m}_{\delta}].
\end{equation*}
By the law of iterated expectation, 
\begin{align*}
    E[\eta^*_{\text{sub}}(\hat{\zeta}) \mid T(\hat{\zeta}) \geq 0, \delta] &= E\{E[\eta^*_{\text{sub}}(\hat{\zeta}) \mid T(\hat{\zeta}) = u, \delta] \mid T(\hat{\zeta}) \geq 0\} \\
    &= \hat{V}_{\delta}A(\hat{\zeta})'[A(\hat{\zeta})\hat{V}_{\delta}A(\hat{\zeta})']^{-1}[E(T(\hat{\zeta}) \mid T(\hat{\zeta}) \geq 0,\delta) - A(\hat{\zeta})\hat{m}_{\delta}]\\
    &\qquad + \hat{m}_{\delta},
\end{align*}
in which $E(T(\hat{\zeta}) \mid T(\hat{\zeta})\geq 0,\delta)$ is the mean of a multivariate normal distribution truncated to the positive orthant,  and can be efficiently computed in R, using \texttt{tmvtnorm::mtmvnorm}. Next, since $\delta$ is treated as a discrete parameter, equation~\eqref{eq:eta_post_mean} reduces to:
\begin{equation*}
    E(\eta_{\text{sub}}(\hat{\zeta}) \mid T(\hat{\zeta}) \geq 0, \hat{\sigma}^2, \hat{\zeta}, y) = \sum_{\ell=1}^D E(\eta^*_{\text{sub}}(\hat{\zeta}) \mid T(\hat{\zeta}) \geq 0, \hat{\sigma}^2, \hat{\zeta}, \delta_\ell, y)p(\delta_\ell \mid \hat{\sigma}^2, \hat{\zeta}, y),
\end{equation*}
using the posterior weights for $\delta_\ell$ defined in equation~\eqref{eq:post_delta}.

\section{Variance estimation}
\label{app:var_est}
We obtain a plugin estimate of $\sigma^2$ by maximizing the likelihood in equation~\eqref{eq:likelihood}. For the optimization done here, $\delta$ and $\zeta$ are not discretized since no integration is necessary. Additionally, only linear equality constraints are imposed on $\eta_{\text{sub}}(\zeta)$ since the inequality constraints in equation~\eqref{eq:cons} are imposed through the prior in our model. 

The estimation of $\sigma^2$ proceeds as follows. Initialize $\sigma^2, \zeta,$ and $\delta$ at a point, $(\sigma^2_{(0)}, \zeta_{(0)}, \delta_{(0)})$, where the subscript denotes the iteration index. For iteration $t = 1, 2, \dots$, repeat the following steps until convergence:
\begin{enumerate}
    \item Given $(\sigma^2_{(t-1)}, \zeta_{(t-1)}, \delta_{(t-1)})$, compute the generalized least-squares estimates of $\alpha, \beta,$ and $\eta_{\text{sub}}(\zeta)$ defined in equations~\eqref{eq:lsalpha}-\eqref{eq:lseta}. Denote these as $(\hat{\alpha}_{(t)}, \hat{\beta}_{(t)}, \hat{\eta}_{\text{sub},(t)}(\zeta_{(t-1)}))$.
    \item Holding $(\hat{\alpha}_{(t)}, \hat{\beta}_{(t)}, \hat{\eta}_{\text{sub}, (t)}(\zeta_{(t-1)}))$ fixed, maximize the likelihood in~\eqref{eq:likelihood} with respect to $(\sigma^2, \zeta, \delta)$ using numerical optimization yielding $(\sigma^2_{(t)}, \zeta_{(t)}, \delta_{(t)})$.
\end{enumerate}
The optimization in Step 2 is performed using the  L-BFGS-B method via \texttt{optim} in R, subject to box constraints $\sigma^2 \in (0.2, \infty), \zeta \in (0, 4]$ and $\delta \in [1, 3]$. Convergence is determined by the default stopping criteria of \texttt{optim} for L-BFGS-B \citep{byrdLimitedMemoryAlgorithm1995}, which terminates when the relative change in the objective function between successive iterations falls below approximately $10^{-9}$, or stop when the iteration limit ($t = 100$) is reached. The resulting estimate, $\hat{\sigma}^2$, is the final iterate $\sigma^2_{(t)}$. Estimates of the remaining parameters obtained during this procedure are not used in subsequent analysis. 

\section{Posterior predictive samples}
\label{app:ppc}
We describe here the procedure for generating posterior predictive samples for the model with likelihood and prior defined in Sections~\ref{sec:likelihood} and~\ref{sec:prior}, respectively.
Recall that $\sigma^2$ is assigned a point-mass empirical Bayes prior fixed at $\sigma^2 = \hat{\sigma}^2$. The posterior can be factorized as
\begin{equation*}
    p(\theta \mid y) \propto p(\eta_{\text{sub}}(\zeta) \mid \alpha, \beta, \hat{\sigma}^2, \zeta, \delta, y) p(\alpha \mid \hat{\sigma}^2, y)p(\beta \mid \hat{\sigma}^2, y)p(\zeta, \delta \mid \hat{\sigma}^2, y),
\end{equation*}
where the conditional distributions of $\alpha$ and $\beta$ are normal, and $\eta_{\text{sub}} \mid \alpha, \beta, \hat{\sigma}^2, \zeta, \delta, y$ follows a truncated multivariate normal distribution (see Appendix~\ref{app:pointests}). 

We sample $(\zeta, \delta)$ over their discrete joint grid using the approximation 
\begin{equation*}
   \hat{p}(\zeta_r, \delta_\ell \mid \hat{\sigma}^2, y) = \frac{p(y \mid \hat{\sigma}^2, \zeta_r, \delta_\ell)p(\delta_\ell)p(\zeta_r)\Delta_{\delta}\Delta_\zeta}{\sum_{r=1}^R \sum_{\ell=1}^D p(y \mid \hat{\sigma}^2, \zeta_r, \delta_\ell)p(\delta_\ell)p(\zeta_r)\Delta_{\delta}\Delta_\zeta},
\end{equation*}
which is computed in the process of obtaining $p(\zeta \mid \hat{\sigma}^2, y)$. A single posterior draw, $\theta^{(t)}$, is obtained sequentially:
\begin{enumerate}
    \item Sample $(\zeta^{(t)}, \delta^{(t)}) \sim \hat{p}(\zeta, \delta \mid \hat{\sigma}^2, y),$
    \item Sample $\alpha^{(t)} \sim p(\alpha \mid \hat{\sigma}^2, y)$ and $\beta^{(t)} \sim p(\beta \mid \hat{\sigma}^2, y)$ (note these do not depend on the samples obtained in Step 1). 
    \item Sample $\eta^{(t)}_{\text{sub}}(\zeta^{(t)}) \sim p(\eta \mid \alpha^{(t)}, \beta^{(t)}, \hat{\sigma}^2, \zeta^{(t)}, \delta^{(t)}, y)$ using the function \texttt{rtmvnorm} in the \texttt{tmvtnorm} package in R with moments defined above, and constraints imposed through $A(\zeta^{(t)})$. The remaining components of $\eta^{(t)}(\zeta^{(t)})$ are then obtained deterministically from the model constraints.
\end{enumerate}
Finally, a posterior predictive draw, $\tilde{y}^{(t)}$, is obtained by sampling from the likelihood conditioned on the posterior draw. That is, $\tilde{Y}^{(t)} \sim p(y \mid \theta^{(t)})$. Repeating this process for $t = 1,...,M$ yields a collection of posterior predictive samples suitable for model checking. 

\subsection{Posterior residuals}
\label{app:residual}
For model assessment, we examine the (standardized) residual curves, defined as wavelength-specific (standardized) residuals obtained by subtracting the posterior predictive mean from the observed data. For a single spectrum, at wavelength index $i$, define the standardized residual as
\begin{equation}
\label{eq:resid_std}
    r_i = \frac{y_i - \frac{1}{M}\sum_{t=1}^M E(\tilde{Y}_i^{(t)} | \theta^{(t)}, y)}{\sqrt{S_{ii}(\hat{\sigma}^2)}},
\end{equation}
where $\theta^{(t)}$ denotes the $t$-th posterior sample, $\tilde{Y}_i^{(t)}$ is the $i$-th entry of $\tilde{Y}^{(t)}$, the posterior predictive random vector corresponding to the $t$-th sample, $E(\tilde{Y}_i^{(t)}| \theta^{(t)}, y)$ denotes the posterior predictive mean flux value at the $i$-th wavelength, $S_{ii}(\hat{\sigma}^2)$ denotes the $i$-th diagonal entry of $S(\hat{\sigma}^2)$, and $M$ is the total the number of posterior samples. The numerator therefore subtracts the posterior predictive mean (approximated by Monte Carlo averaging) from the observed flux.

One could alternatively use posterior predictive draws, $\tilde{y}_j^{(t)}, \; t=1,\dots,M$, directly in place of the conditional expectations. However, doing so would introduce additional Monte Carlo noise into the residuals and obscure assessment of the posterior predictive mean fit.

\subsection{Posterior predictive p-values}
For model assessment, we also consider posterior predictive p-values (PPPs). The specific PPPs used in Appendix~\ref{app:model_check} are described here. We distinguish between \emph{global} PPPs, which produce a single scalar summary for each spectrum, and \emph{local} PPPs which produce a vector of values indexed by wavelength.

We first recall the definition of a global PPP. Given posterior draws $\theta^{(t)},$ for $t = 1,\dots,M$, and the associated posterior predictive samples, $\tilde{y}^{(t)}$, let $T(y, \theta)$ denote a discrepancy function which is some measure of model fit (e.g., mean squared error), which may depend on both the data and the parameters. The global PPP is then estimated via Monte Carlo as 
\begin{equation*}
    \widehat{PPP} = \frac{1}{T}\sum_{t = 1}^T \mathbb{I}_{\{T(\tilde{y}^{(t)}, \theta^{(t)}) \geq T(y, \theta^{(t)})\}},
\end{equation*}
where $\mathbb{I}_{\{\cdot\}}$ denotes the indicator function, and $y$ denotes the observed flux vector. We consider three global discrepancies: mean squared error ($T_{\text{MSE}}$), the maximum standardized residual ($T_{\text{Max}}$), and the standardized median absolute deviation from the median ($T_{\text{MAD}}$). Formally, they are defined as
\begin{align}
    T_{\text{MSE}}(y, \theta) &= \frac{1}{N}\sum_{i=1}^N \frac{(y_{i} - E(\tilde{Y}_i \mid \theta, y))^2}{S_{ii}(\hat{\sigma}^2)}, \label{eq:tmse}\\
    T_{\text{Max}}(y, \theta) &= \max_i\left\{\frac{|y_{i} - E(\tilde{Y}_i \mid \theta, y)|}{\sqrt{S_{ii}(\hat{\sigma}^2)}}\right\}, \label{eq:tax}\\
    T_{\text{MAD}}(y, \theta) &= \text{med}_i\left\{\frac{|y_{i} - \text{med}(\tilde{Y}_i \mid \theta, y)|}{\sqrt{S_{jj}(\hat{\sigma}^2)}}\right\}, \label{eq:tmad}
\end{align}
where $E(\tilde{Y}_i \mid \theta, y)$ and $\text{med}(\tilde{Y}_i \mid \theta, y)$ represent the posterior predictive mean and median flux value at the $i$-th wavelength, respectively. Next, we consider a local PPP based on squared standardized residuals, where the $j$-th local PPP is given by
\begin{equation}
\label{eq:local_ppp}
    T_{\text{local}, i}(y, \theta) = \frac{(y_i - E(\tilde{Y}_i \mid \theta, y))^2}{S_{ii}(\hat{\sigma}^2)}.
\end{equation}
That is, we simply compute the $i$-th entry in the summation defining $T_{\text{MSE}}(y, \theta)$, but do not average across wavelength. These local PPPs are useful in that they allow us to identify indices $i$ for which the local PPPs are small. For example, if $i = 4, 5, 6, 7, 8$ all lead to small PPPs, then we expect some structural model misfit for the corresponding wavelengths.

\begin{figure}[t]
    \centering
    \includegraphics[width=\linewidth]{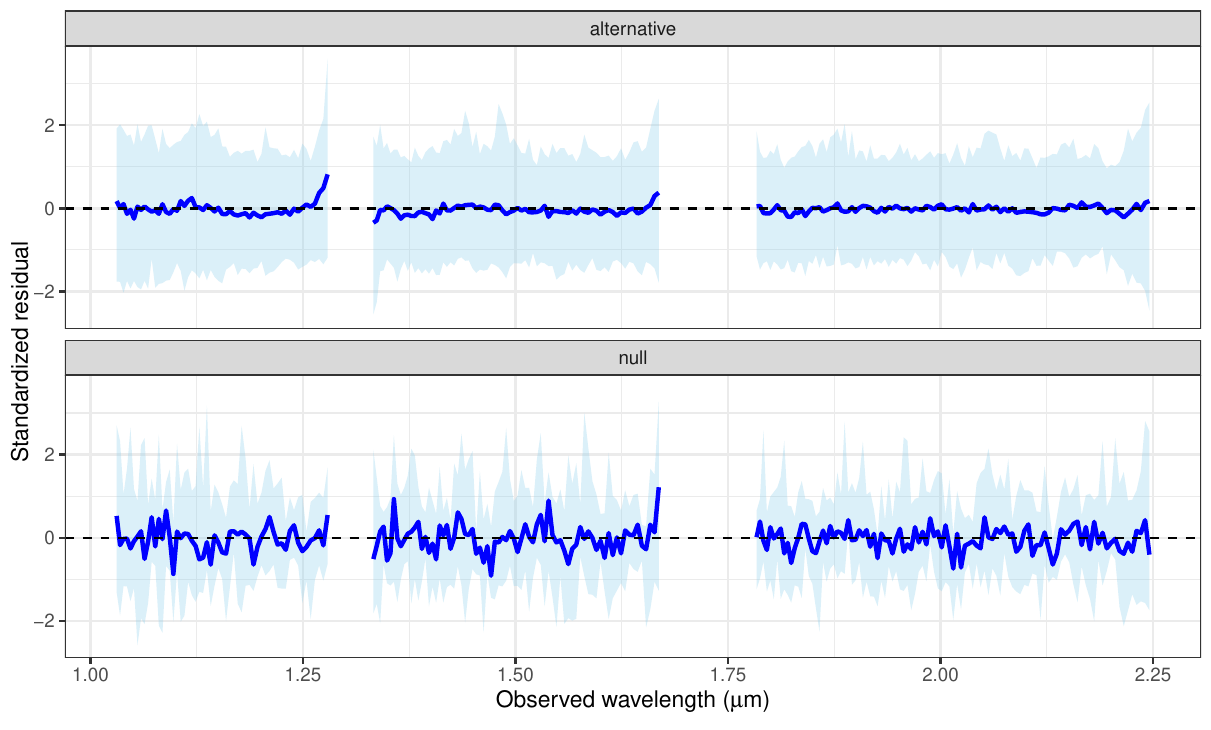}
    \caption{Aggregated standardized residual curves for the 213 labeled spectra, separated according to log posterior odds model selection into 203 alternative cases (top) and 10 null cases (bottom). The dark blue curve shows the pointwise median of the standardized residuals across spectra, and the light blue band represents the 2.5\% and 97.5\% pointwise quantiles. The greater variability in the null panel reflects the smaller number of spectra classified as null.}
    \label{fig:resid_curve_labeled}
\end{figure}

\section{Model checking}
\label{app:model_check}
This appendix presents graphical and numeric diagnostics to assess model fit. These includes residual curves based on posterior samples posterior predictive p-value summaries (as defined in Appendix~\ref{app:ppc}). Model checking for spectral data is somewhat challenging duet to the presence of sharp emission features, which generate extreme observations. Residual-based diagnostics must therefore be interpreted with care, as large residuals may reflect emission lines that are slightly underfit rather than significant model misfit overall. Below, we describe diagnostics designed to characterize model fit in this setting.

\begin{figure}[ht]
    \centering
    \includegraphics[width=\linewidth]{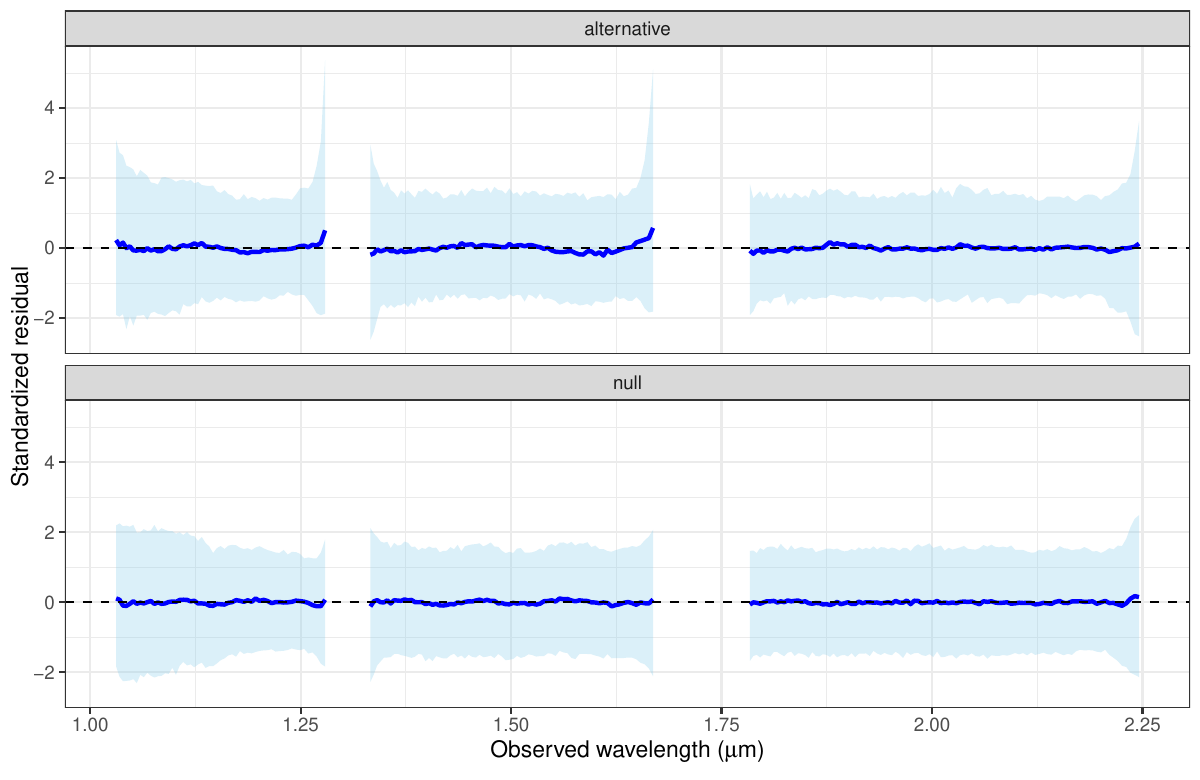}
    \caption{Aggregated standardized residual curves for the 2800 unlabeled spectra, separated according to log posterior odds model selection into 937 alternative cases (top) and 1863 null cases (bottom). The dark blue curve shows the pointwise median of the standardized residuals across spectra, and the light blue band represents the 2.5\% and 97.5\% pointwise quantiles. The greater variability in the null panel reflects the smaller number of spectra classified as null.}
    \label{fig:resid_curve_unlabeled}
\end{figure}

\subsection{Residual curves: 213 labeled spectra}
We first assess the mean fit (background and signal components) across the 213 labeled spectra. For each spectrum, the per-wavelength residuals $r_i$, $i = 1, \dots, N$, defined in Appendix~\ref{app:residual}, are indexed by wavelength and therefore form a standardized residual curve over the $N$ observed wavelength values. Model fit is evaluated after model selection. Specifically, for each spectrum we select either the null model, $M_0$, or the alternative model, $M_1$, based on the log posterior odds. Residual curves are then aggregated separately within each selected model. Figure~\ref{fig:resid_curve_labeled} displays the aggregated curves as functions of wavelength across the 203 spectra classified as $M_1$ and the 10 spectra classified as $M_0$. The posterior predictive means in equation~\eqref{eq:resid_std} are approximated using $M = 1000$ posterior draws for each spectrum.

For both the null and alternative cases, ideal model fit would produce residual curves whose median fluctuates around zero, with the 95\% pointwise quantile bands largely contained within $\pm 2$. Because the aggregation is based on only 203 spectra for $M_1$ and 10 spectra for $M_0$, some variability around zero is expected. The null panel exhibits substantially more variability due to the small sample size. 

The primary diagnostic concern is systematic deviation from zero. Such behavior is most evident near the boundaries of the observed wavelength range. In particular, near the edges and detector gaps the median residual curve shows noticeable deviations from zero, especially under the alternative model. This reflects the limited information available to constrain the background and emission line model components in these regions due to being near a boundary. Away from detector gaps, the residual curves generally behave like constant functions centered around zero with some random fluctuation.

\subsection{Residual curves: 2800 unlabeled spectra}
Figure~\ref{fig:resid_curve_unlabeled} shows aggregated residual curves for the 2800 unlabeled spectra, separated according to log posterior odds model selection into 937 alternative cases and 1863 null cases. Because this dataset contains substantially more spectra classified under both models, the aggregated residual curves are smoother and more stable than those observed for the labeled case. For spectra with detected emission lines, similar edge effects are observed near detector gaps, consistent with the behavior seen in the labeled data. These boundary effects are considerably less pronounced in the null case. 

\begin{figure}[ht]
    \centering
    \includegraphics[width=\linewidth]{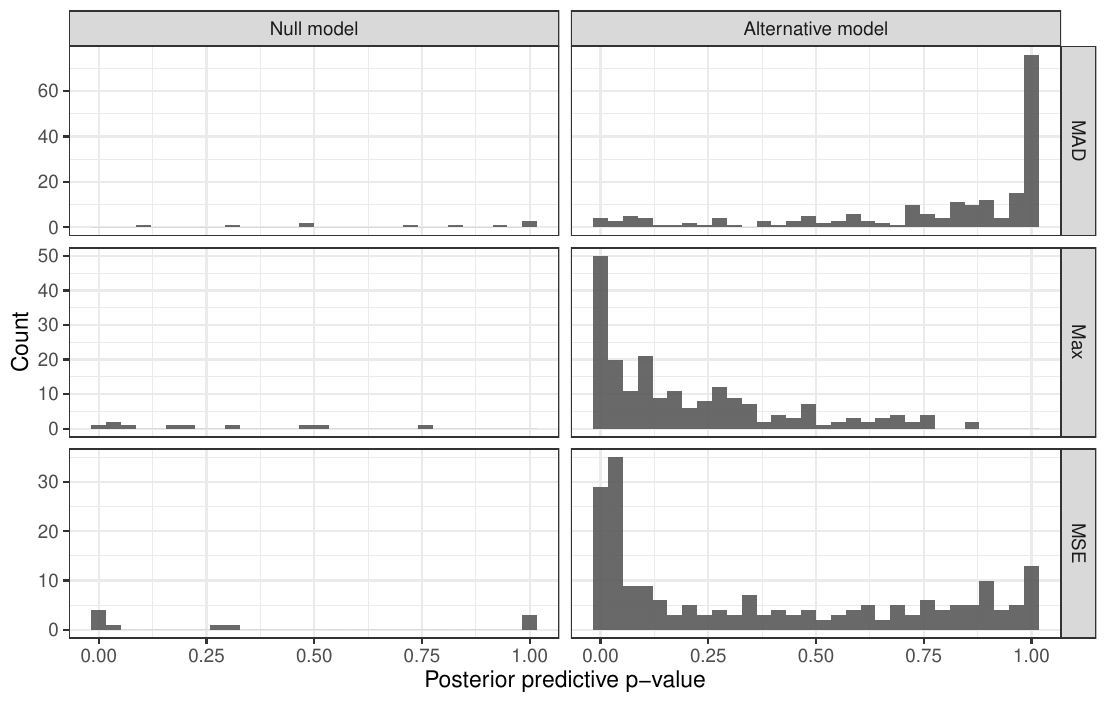}
    \caption{Histograms of posterior predictive p-values for the 213 labeled spectra, computed using the $T_{\text{MAD}}$, $T_{\text{Max}}$, and $T_{\text{MSE}}$ discrepancies in \eqref{eq:tmse}–\eqref{eq:tmad} (top to bottom). Spectra are first classified according to the log posterior odds model selection criterion, and goodness-of-fit is then evaluated within the selected model. The left column corresponds to spectra classified as null and the right column to those classified as alternative. The null histograms are sparse due to the small number of spectra selected as null in the labeled set.}
    \label{fig:ppp_hists_labeled}
\end{figure} 

In both panels, the median residual curves remain close to zero across most wavelengths, indicating that, after model selection, the selected model captures the overall mean structure of the spectra reasonably well within each group. Patterns in residual variability are evident, however. Most notably, over the wavelength range $[1.03\,\mu\mathrm{m}, 1.27\,\mu\mathrm{m}]$, the spread of the standardized residuals decreases monotonically in both the null and alternative cases, a pattern not observed elsewhere. This suggests residual heteroskedasticity in the observed data beyond what is captured by the reported measurement errors. 

Because the variance model takes the form $\sigma^2 I_N + C$ as in equation~\eqref{eq:cov}, heteroskedasticity enters only through the reported measurement errors, while the estimated component $\sigma^2$ is global. Consequently, wavelength-dependent miscalibration of the measurement errors is not corrected by the model and persists in the posterior predictive distribution. The contraction of the standardized residual bands over $[1.03\,\mu\mathrm{m}, 1.27\,\mu\mathrm{m}]$ indicates that variability may be slightly overestimated there. This form of misspecification is less concerning than systematic underestimation of variance, as it generally produces more conservative inference, including weaker evidence for line detection and wider redshift HPD sets. 

\subsection{Posterior predictive checks: 213 labeled spectra}
Figure~\ref{fig:ppp_hists_labeled} shows histograms of global posterior predictive p-values (PPPs) for the 213 labeled spectra. A substantial portion of spectra exhibit extreme PPP values (near 0 or 1), which at first glance suggests poor model overall fit. However, these discrepancies are based on global summaries and are sensitive to isolated outliers and the quality of the variance estimation.

\begin{figure}[ht]
    \centering
    \includegraphics[width = 0.95\textwidth, height = 0.34\textheight]{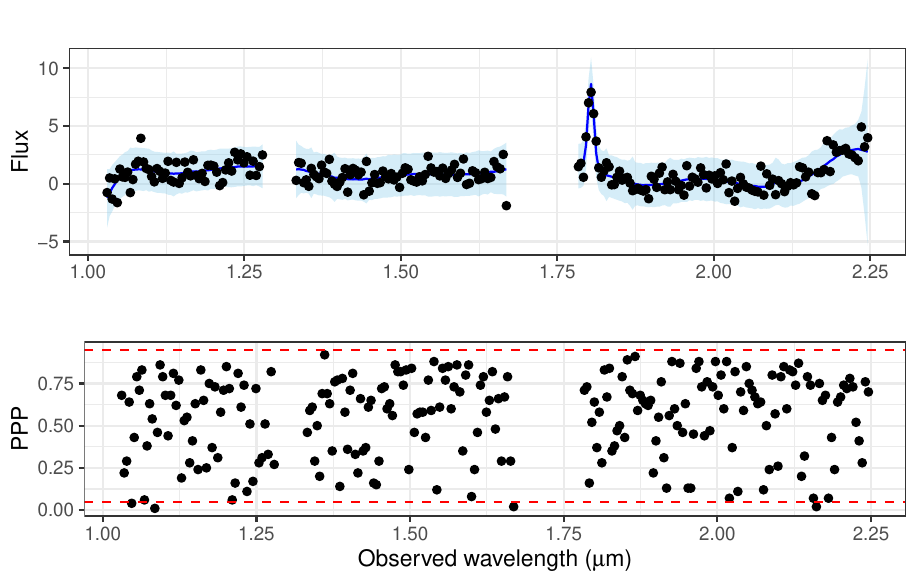}
    \caption{The top panel shows the observed data for spectrum 00004 (black dots) together with the pointwise median and 2.5\% and 97.5\% posterior predictive quantiles (dark and light blue). The bottom panel shows the corresponding pointwise PPPs based on $T_{\text{local}}$ in Appendix~\ref{app:residual}, with dashed red lines at 0.05 and 0.95. A few wavelengths have PPPs below 0.05, but none are consecutive.}
    \label{fig:spec_01_plus_ppp}
\end{figure}

\begin{figure}[ht]
    \centering
    \includegraphics[width = 0.95\textwidth, height = 0.35\textheight]{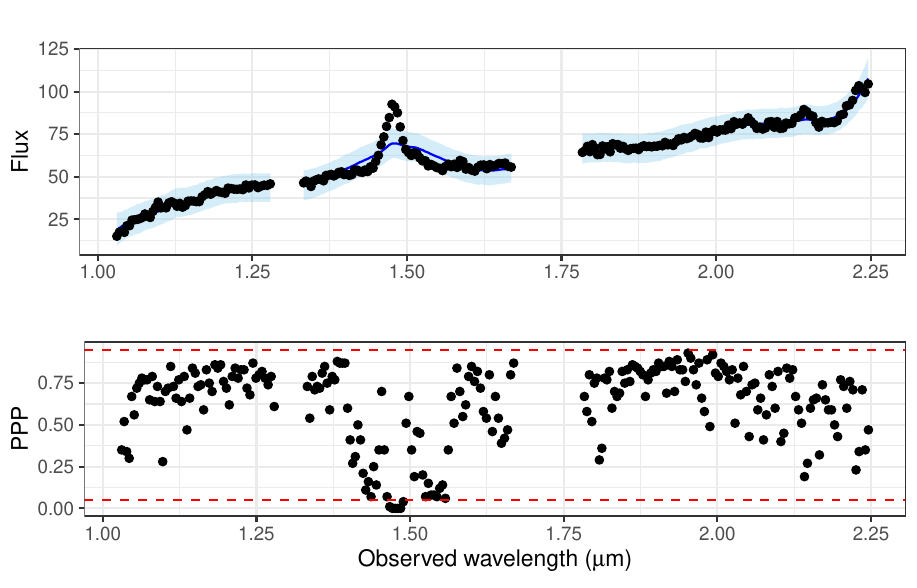}
    \caption{The top panel shows the observed data for spectrum 00201 (black dots) together with the pointwise median and 2.5\% and 97.5\% posterior predictive quantiles (dark and light blue). The prominent bump is not captured by the posterior predictive intervals. The bottom panel shows the corresponding pointwise PPPs based on $T_{\text{local}}$ in \eqref{eq:local_ppp}, with dashed red lines at 0.05 and 0.95. A sequence of PPPs below 0.05 aligns with this lack of fit, and the PPPs are shifted above 0.5, indicating inflated posterior predictive variance at each wavelength.}
    \label{fig:spec_14_plus_ppp}
\end{figure}

\begin{figure}[ht]
    \centering
    \includegraphics[width=\linewidth]{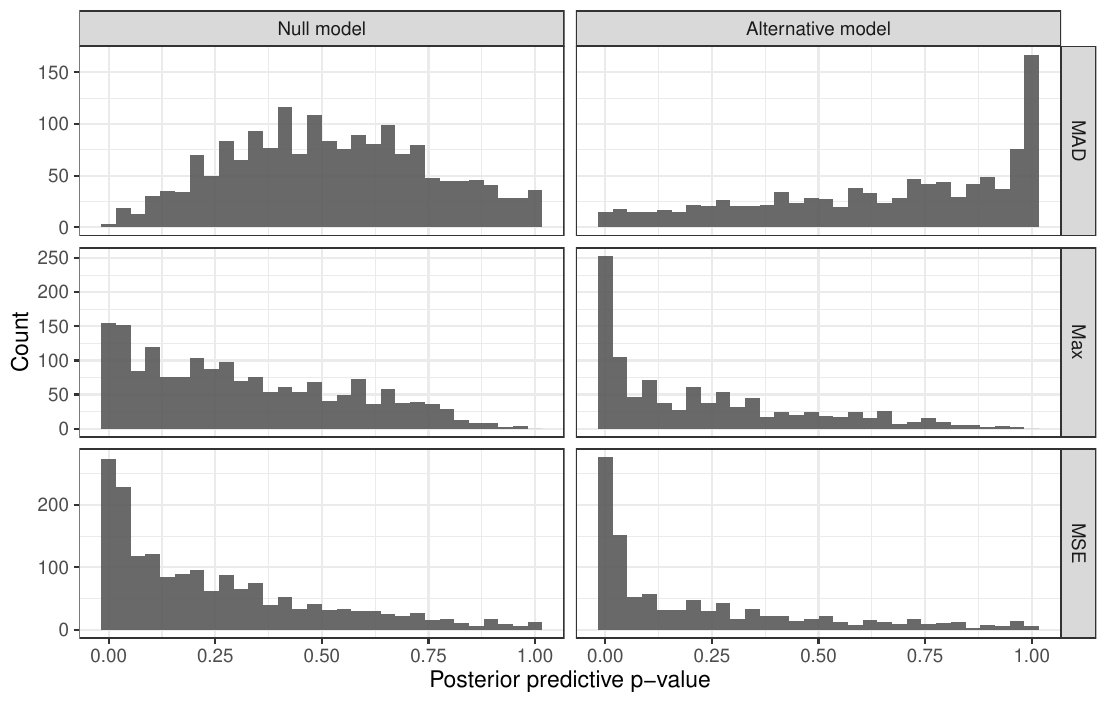}
    \caption{Histograms of posterior predictive p-values for the 2800 unlabeled spectra, computed using the $T_{\text{MAD}}$, $T_{\text{Max}}$, and $T_{\text{MSE}}$ discrepancies in \eqref{eq:tmse}–\eqref{eq:tmad} (top to bottom). Spectra are first classified according to the log posterior odds model selection criterion, and goodness-of-fit is then evaluated within the selected model. The left column corresponds to spectra classified as null and the right column to those classified as alternative.}
    \label{fig:ppp_hists_unlabeled}
\end{figure}

In particular, PPPs near zero typically arise when the posterior predictive distribution fails to cover the observed flux at one or more wavelengths. This commonly occurs in cases of slight underfitting of strong emission lines. Because the global discrepancies aggregate across wavelengths, a single large residual can drive the PPP close to zero. For example, spectrum 00033 in Figure~\ref{fig:spec_04} of the main text has a $T_{\text{MSE}}$-based PPP of zero, even though the overall posterior predictive fit is satisfactory. Conversely, PPP values near one indicate that the predictive discrepancies are systematically larger than the observed discrepancy. This behavior is often consistent with predictive overdispersion due, for example, to inflated variance estimates. In this setting, such misspecification is less concerning, as it primarily leads to conservative inference as noted previously. The presence of strong emission lines itself can inflate variance estimates, making this behavior unsurprising.

Because the global PPPs are sensitive to isolated deviations, we focus on structural model failures using local PPPs based on \eqref{eq:local_ppp} computed at each wavelength. These allow identification of systematic discrepancies across contiguous wavelength regions, such as underfitting of emission lines or background misspecification. To flag such behavior, we identify spectra exhibiting multiple consecutive local PPPs near zero (e.g., we flag any spectra with four consecutive PPPs below 0.05). 

The top panel of Figure~\ref{fig:spec_01_plus_ppp} shows the posterior predictive distribution for spectrum 00004, which appears to fit well according to visual assessment as discussed in the main text. The bottom panel of the figure shows the corresponding sequence of local PPPs across wavelength, with dashed red lines marking extreme range cutoffs of 0.05 and 0.95 (chosen somewhat arbitrarily). Although a few wavelengths yield small PPP values, they are isolated rather than consecutive. The corresponding global PPPs ($T_{\text{MAD}}, T_{\text{Max}}$, and $T_{\text{MSE}}$) are 0.97, 0.59, and 1, respectively, for reference. 

In contrast, Figure~\ref{fig:spec_14_plus_ppp} shows the same two plots for spectrum 00201, which exhibits a prominent emission line feature not captured by the posterior predictive distribution. This corresponds to a sequence of consecutive small local PPPs aligned with the structural discrepancy. The global PPPs for this spectrum are 1, 0, and 0.84 for $T_{\text{MAD}}$, $T_{\text{Max}}$, and $T_{\text{MSE}}$, respectively, reflecting both localized underfitting and variance inflation.

\begin{figure}[ht]
    \centering
    \includegraphics[width=\linewidth]{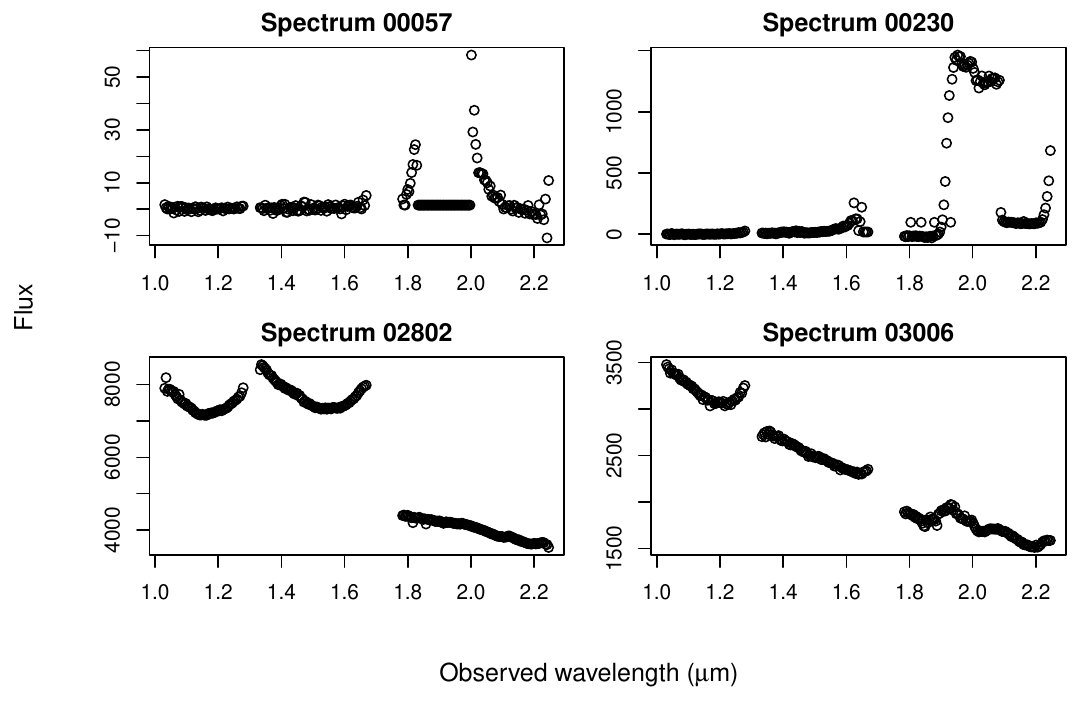}
    \caption{Plots of the observed flux values versus wavelength for four low quality spectra out of the unlabeled set of 2800. All four of these spectra are flagged as having poor model fit, and are removed from the main results due to obvious errors in the detection and data recording process.}
    \label{fig:garbage_spectra}
\end{figure}

Taken together, these examples illustrate that extreme global PPP values often reflect localized deviations rather than widespread model failure. Local PPPs provide a more informative diagnostic by identifying contiguous regions of misfit. In total, for the 213 labeled spectra, using a identification rule of five consecutive local PPPs less than or equal to 0.05, there are nine spectra identified for follow-up due to potentially poor model fit. Note that while the posterior predictive fit is poor, across the nine flagged spectra, the largest absolute difference between the marginal MAP redshift estimate and the reported redshift from astronomers is 0.00187. That is, these fits do not lead to pathological redshift fitting as they still identify the locations of emission lines even if heights of emission lines are underfit, or the variance is poorly estimated. 

\subsection{Posterior predictive checks: 2800 unlabeled spectra}
Figure~\ref{fig:ppp_hists_unlabeled} shows histograms of global posterior predictive p-values (PPPs) for the 2800 unlabeled spectra. For spectra selected as containing emission lines, the PPP behavior closely mirrors that observed in the labeled sample. Under the null model, however, the substantially larger number of spectra allows a clearer assessment of discrepancy behavior. The $T_{\text{MSE}}$ and $T_{\text{MSE}}$ discrepancies remain sensitive to isolated extreme residuals (as they should), often producing PPPs near 0 or 1. In contrast, the $T_{\text{MAD}}$ discrepancy is more robust to outliers. Under the null model, the $T_{\text{MAD}}$-based PPPs are concentrated around 0.5 with an approximately symmetric decline toward 0 and 1, consistent with PPP behavior under ``adequate'' model fit. 

Posterior predictive p-values are known to exhibit conservatism and do not follow a uniform distribution under the null. Nevertheless, the concentration of $T_{\text{MAD}}$-based PPPs near 0.5 in the null case suggests that the background and noise components of the model are generally well calibrated when no emission lines are present. The more extreme values observed for the $T_{\text{MSE}}$ and $T_{\text{Max}}$ discrepancies primarily reflect sensitivity to localized extremes in the observed spectra.

Using a local PPP summary instead, we flag spectra that contain at least one sequence of five consecutive local PPPs below 0.05. This threshold is somewhat arbitrary, but we saw in the high-quality data setting that it meaningfully flagged poor structural model fit. This criterion identifies 85 of the 2800 spectra as exhibiting poor model fit. Figure~\ref{fig:garbage_spectra} displays four representative examples.

The flagged spectra exhibit several recurring data quality issues. Spectrum 00057 contains an extended stretch of flux values fixed at zero, likely reflecting truncation or missing values. Spectrum 00230 shows an abrupt jump in flux values, increasing from values between 0 and 100 to values near 1500, consistent with detector or data-processing artifacts. Spectra 02802 and 03006 contain extremely large flux values (exceeding 8000 and 3000, respectively), likely caused by contamination from a nearby bright source rather than the target galaxy.

These examples illustrate structural data errors that would likely compromise inference under any modeling procedure. We therefore remove all 85 flagged spectra from the unlabeled analysis. While this filtering rule may exclude a small number of spectra with reliable data but imperfect model fit, in general the observed misfit appears to be driven primarily by data corruption. In this setting, removing suspect spectra is preferable to retaining observations that would yield unreliable redshift estimates.

\section{Prior hyperparameter sensitivity analysis}
\label{app:sens}

In the sensitivity analysis that follows, each subsection examines variation in the hyperparameters of a single model component, while holding all other prior hyperparameters fixed at the values described in Section~\ref{sec:prior} and used in the main analysis. For example, when assessing sensitivity to the signal intensity prior, the intercept prior, background specification (including the number of components and prior variances), line width prior, and redshift prior are held fixed at their values used in the main analysis. Sensitivity results are conducted using only the 213 high-quality spectra.

For each prior hyperparameter setting, we report the following summaries: (i) histograms of the log posterior odds; (ii) a table of rejection rates based on the log posterior odds, and false positive rates on simulated data; (iii) histograms of the resulting redshift marginal MAP estimates; and (iv) boxplots of total HPD widths. 

To estimate false positive rates (FPRs) under each hyperparameter configuration, we generate synthetic null datasets as follows. For each of the 213 labeled spectra, we first fit the proposed model under the null hypothesis using the baseline priors from the main analysis. Using the resulting posterior estimates under the null model, we generate one synthetic dataset from the corresponding null-model likelihood. This produces 213 synthetic null spectra, one for each original JWST spectrum.

The synthetic null data are generated using the baseline prior specification rather than re-fitting under each alternative hyperparameter configuration. This ensures that the null data-generating mechanism remains fixed across sensitivity settings. For each prior hyperparameter configuration under consideration, we then reapply the full detection procedure to the 213 synthetic null spectra and record the proportion of cases in which the alternative model is selected. This proportion serves as an empirical estimate of the FPR under that prior specification. 

\subsection{Alpha hyperparameter}
The results in the main paper were replicated with the priors 
$\alpha \sim N(5, 10^2)$ and $\alpha \sim N(0, 100^2)$. The results were numerically identical to the main analysis and are thus omitted.

\begin{figure}[ht]
    \centering
    \includegraphics[width=\linewidth]{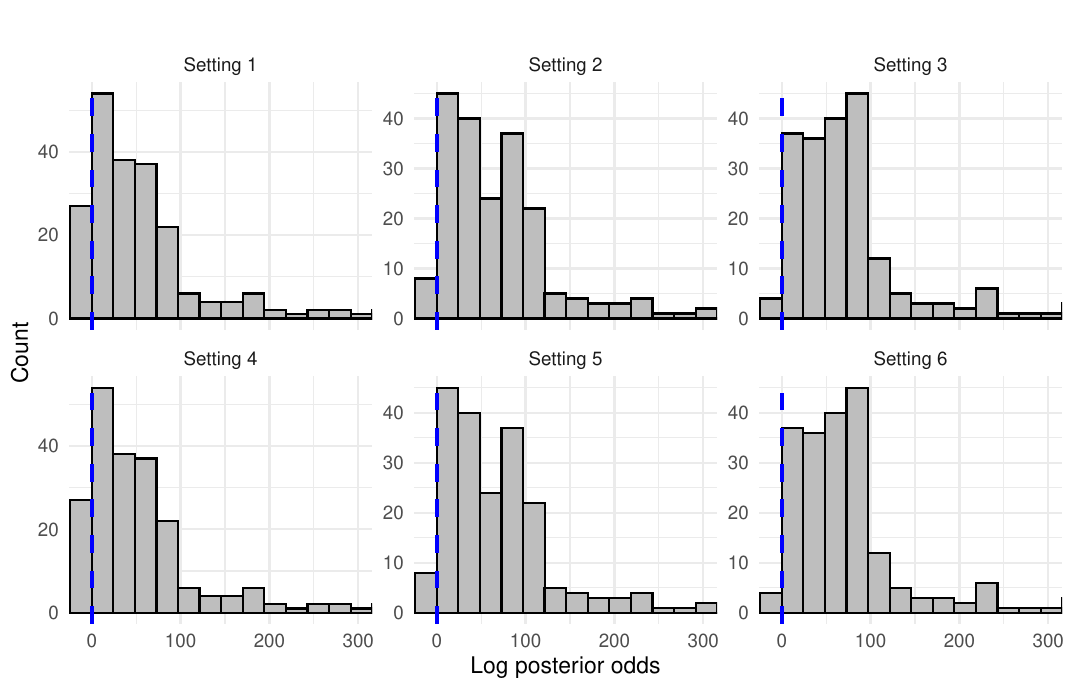}
    \caption{Each histogram depicts the log posterior odds for a different hyperparameter setting for the number of background basis functions and beta coefficient prior variances. A small number of large log posterior odds values are cutoff from each plot. The dashed blue line at zero indicates the cutoff threshold.}
    \label{fig:log_PO_hists_background}
\end{figure}

\begin{table}[ht]
\centering
\begin{tabular}{lcc}
\hline
Setting & Rejection rate (Signal) & FPR (Null) \\
\hline
1 & 0.873 & 0.019 \\
2 & 0.962 & 0.019 \\
3 & 0.981 & 0.901 \\
4 & 0.873 & 0.019 \\
5 & 0.962 & 0.019 \\
6 & 0.981 & 0.901 \\
\hline
\end{tabular}
\vspace{6pt}
\caption{Rejection probabilities on JWST data labeled as containing a signal by astronomers and the false positive rate (FPR) on simulated null data. Sample size of $n = 213$ per setting; maximum Monte Carlo standard error $\le 0.023$.}
\label{table:sens_background}
\end{table}

\subsection{Beta hyperparameter and number of bases}
\label{app:bg}
To assess the sensitivity to the flexibility of the background function, we vary both the number of basis functions, $J$, and the prior variance for each $\beta_j$. Specifically, we consider the following cases:
\begin{enumerate}
    \item $J = 30$ with $b_1^2 = 30^2$
    \item $J = 15$ with $b_1^2 = 30^2$
    \item $J = 5$ with $b_1^2 = 30^2$
    \item $J = 30$ with $b_1^2 = 10^2$
    \item $J = 15$ with $b_1^2 = 10^2$
    \item $J = 5$ with $b_1^2 = 10^2$.
\end{enumerate}
Setting 2 ($J = 15$, $b_1 = 30$) corresponds to the prior specification used in the main analysis.

\begin{figure}[ht]
    \centering
    \includegraphics[width=\linewidth]{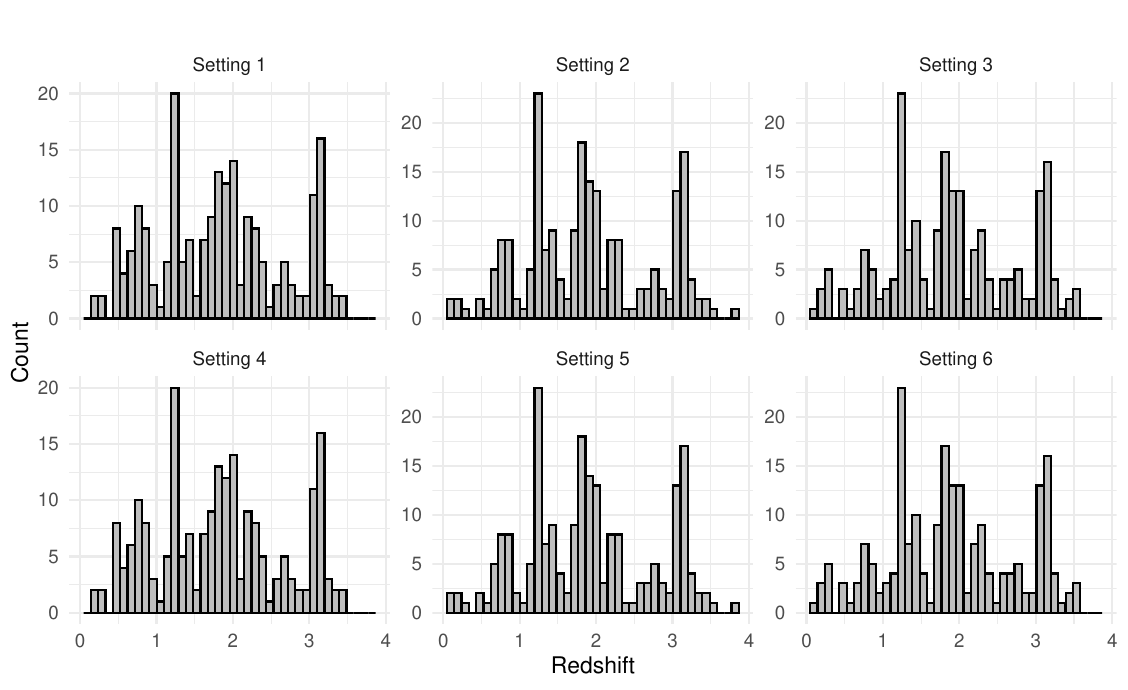}
    \caption{Each histogram depicts the distribution of marginal MAP redshift estimates for a different hyperparameter setting for the number of background basis functions and beta coefficient prior variances.}
    \label{fig:redshift_hists_background}
\end{figure}

\subsubsection{Impact on log posterior odds and FPRs}
Figure~\ref{fig:log_PO_hists_background} displays six histograms of the log posterior odds corresponding to the hyperparameter settings described above. In the top row of the figure, moving from left to right (with fixed prior variance $b_1^2 = 30^2$), decreasing the number of basis functions, $J$, shifts the distribution of the log posterior odds towards positive values. This shift is practically meaningful, as it results in a larger number of spectra being identified as containing emission lines. The bottom row exhibits an identical pattern for $b_1^2 = 10^2$. That is, changing the prior variance from $30^2$ to $10^2$ has no effect on the log posterior odds shown here.

Table~\ref{table:sens_background} reports the rejection rates for the 213 JWST spectra and the corresponding false positive rates (FPRs) based on 213 simulated null spectra. The results again show that sensitivity to the background specification is dictated by the number of basis functions $J$. When $J = 5$ basis functions are used, the FPR increases dramatically to 0.901, indicating severe anti-conservative behavior arising from an inflexible background model. The accompanying increase in rejection rate is therefore misleading. In contrast, for $J = 15$ and $J = 30$, the FPR remains low and stable, and is unaffected by the choice of prior variance in this range. Although the FPR is identical for $J = 15$ and $J = 30$, the $J = 30$ configuration yields slightly lower rejection rates overall, showing that the detection power is mildly sensitive to the degree of background flexibility (as was observed with the log posterior odds distributions). 

\subsubsection{Impact on redshift distributions}
Next, we examine the sensitivity of redshift estimation to changes in background hyperparameters. Figure~\ref{fig:redshift_hists_background} presents six histograms of marginal MAP redshift estimates for all 213 spectra. In contrast to the main analysis, we do not restrict attention to spectra identified as containing emission lines, in order to assess the overall impact of the hyperparameter settings on redshift variability.

Once again, the results are effectively identical across the prior variance choices, and depend primarily on $J$. Across $J = 5, 10, 15$, the redshift distributions are broadly similar, though some differences are visible. For example, the $J = 30$ configuration exhibits a slightly higher concentration of low-redshift estimates relative to the other cases. 

\begin{figure}[ht]
    \centering
    \includegraphics[width=\linewidth, height = 0.35\textheight]{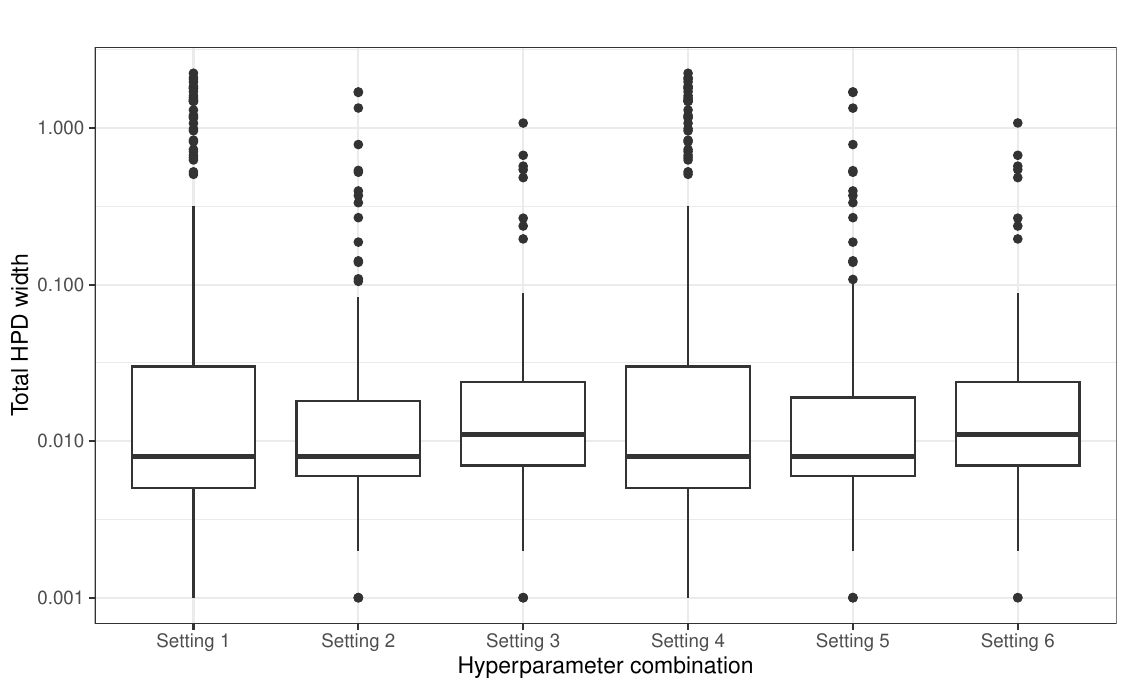}
    \caption{Each histogram depicts the total HPD width (on the log scale) for a different hyperparameter setting for the number of background basis functions and beta coefficient prior variances.}
    \label{fig:hpd_widths_background}
\end{figure}

\begin{figure}[ht]
    \centering
    \includegraphics[width=\linewidth]{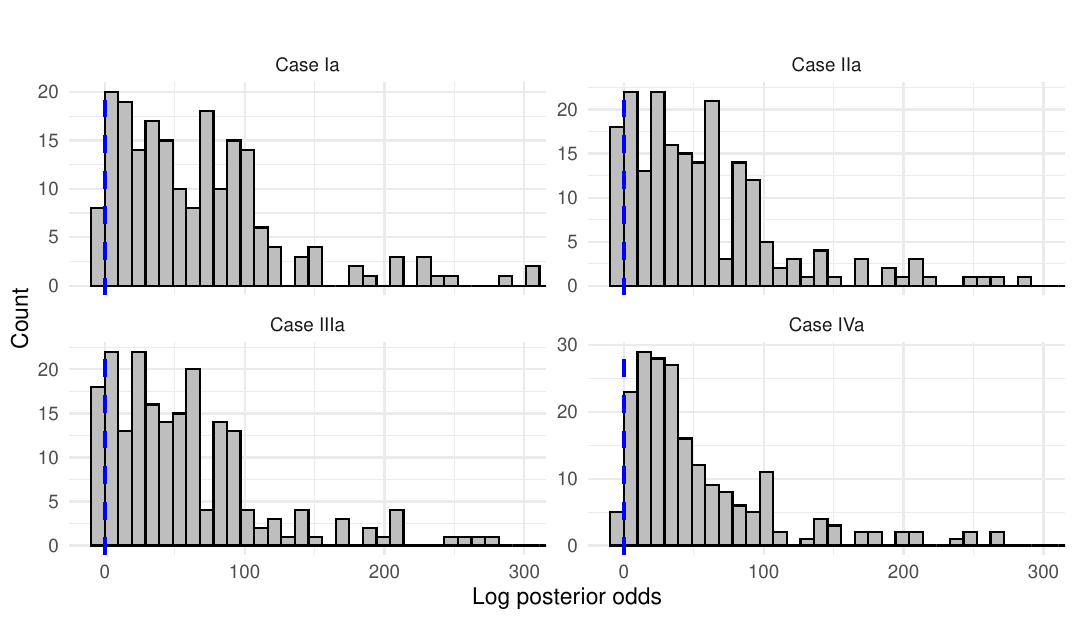}
    \caption{Each histogram depicts the log posterior odds for a different hyperparameter setting for the signal intensities. A small number of large log posterior odds values are cutoff from each plot. The dashed blue line at zero indicates the cutoff threshold.}
    \label{fig:log_PO_hists_etas}
\end{figure}

\begin{table}[ht]
\centering
\begin{tabular}{lcc}
\hline
Setting & Rejection rate (Signal) & FPR (Null) \\
\hline
Ia & 0.962 & 0.019 \\
IIa & 0.915 & 0.005 \\
IIIa & 0.915 & 0.005 \\
IVa & 0.977 & 0.164 \\
\hline
\end{tabular}
\vspace{6pt}
\caption{Rejection probabilities on JWST data labeled as containing a signal by astronomers and the false positive rate (FPR) on simulated null data. Sample size of $n = 213$ per setting; maximum Monte Carlo standard error $\le 0.025$.}
\label{table:sens_etas}
\end{table}

\subsubsection{Impact on redshift HPD set widths}
Finally, Figure~\ref{fig:hpd_widths_background} shows boxplots of the total 99.865\% HPD widths for redshift under each hyperparameter configuration (displayed on the log scale). As before, results are identical across prior variance settings and depend only on $J$. The median HPD width is nearly identical for $J = 15$ and $J = 30$, with a modest increase when $J = 5$. Overall, while the HPD widths exhibit some variation across hyperparameter settings, they are less sensitive than the false positive rates.

\subsection{Signal intensity hyperparameters}
\label{app:sig}
We next assess the sensitivity of the results to the signal intensity hyperparameters, which prove to be the most influential component of the prior specification. We consider the following cases:
\begin{itemize}
    \item Case Ia - Science informed prior (used in main text): 
        \begin{align*}
            a_2 &= (0.02, 0.02, 0.0092, 0.02, 0.2, 0.596, 0.2, 0.02, 0.02, 0.0271, 0.02, 0.05, 0.02)',\\
            b_2 &= (0.01, 0.01, 0.01, 0.01, 1, 1, 1, 0.01, 0.01, 0.01, 0.01, 0.01, 0.01)'
        \end{align*}
    \item Case IIa - Prior with scientific informed mean and large equal standard deviations:
        \begin{align*}
            a_2 &= (0.02, 0.02, 0.0092, 0.02, 0.2, 0.596, 0.2, 0.02, 0.02, 0.0271, 0.02, 0.05, 0.02)',\\
            b_2 &= (1, 1, 1, 1, 1, 1, 1, 1, 1, 1, 1, 1, 1)'
        \end{align*}
    \item Case IIIa - Priors with equal means and large equal standard deviations:
        \begin{align*}
            a_2 &= (0.02, 0.02, 0.02, 0.02, 0.02, 0.02, 0.02, 0.02, 0.02, 0.02, 0.02, 0.02, 0.02)',\\
            b_2 &= (1, 1, 1, 1, 1, 1, 1, 1, 1, 1, 1, 1, 1)'
        \end{align*}   
    \item Case IVa - Priors with equal means and small equal standard deviations:
        \begin{align*}
            a_2 &= (0.02, 0.02, 0.02, 0.02, 0.02, 0.02, 0.02, 0.02, 0.02, 0.02, 0.02, 0.02, 0.02)',\\
            b_2 &= (0.01, 0.01, 0.01, 0.01, 0.01, 0.01, 0.01, 0.01, 0.01, 0.01, 0.01, 0.01, 0.01)'
        \end{align*}   
\end{itemize}

\subsubsection{Impact on log posterior odds and FPRs}
Figure~\ref{fig:log_PO_hists_etas} shows histograms of the log posterior odds under the four prior settings described in the previous section. Cases IIa and IIIa are noticeably more conservative than the science-informed prior (Case Ia). Both impose larger and equal prior variances across emission lines, increasing flexibility in the alternative model and thereby introducing a larger penalty in the marginal density of $y$ under the alternative model. As a result, the log posterior odds shift downward relative to Case Ia. Case IVa, in contrast, is the most liberal configuration, identifying only five spectra as null cases. However, Table~\ref{table:sens_etas} shows that the increased detection rate comes at a cost: the FPR is 0.164, compared to 0.019, 0.005, and 0.005 for Cases Ia-IIIa, respectively. The tight prior constraints in Case IVa reduce the effective penalty for the alternative model, making spurious detections more likely. Overall, while many spectra yield consistent decisions across prior choices, detection rates vary meaningfully across configurations. Increasing prior variance generally leads to more conservative behavior, whereas overly tight priors inflate false positives. It is reassuring to see that the science-informed prior, chosen \emph{a priori} based on expert belief, balances this trade-off.

\begin{figure}[ht]
    \centering
    \includegraphics[width=\linewidth]{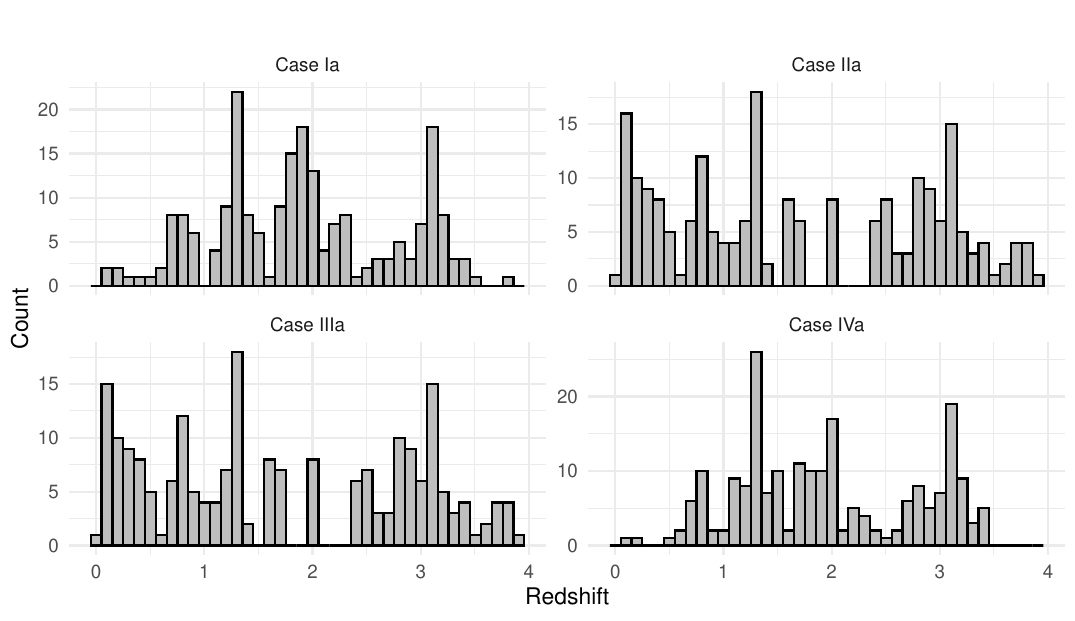}
    \caption{Each histogram depicts the distribution of marginal MAP redshift estimates for a different hyperparameter setting for the signal intensities.}
    \label{fig:redshift_hists_etas}
\end{figure}

\begin{figure}[ht]
    \centering
    \includegraphics[width=\linewidth]{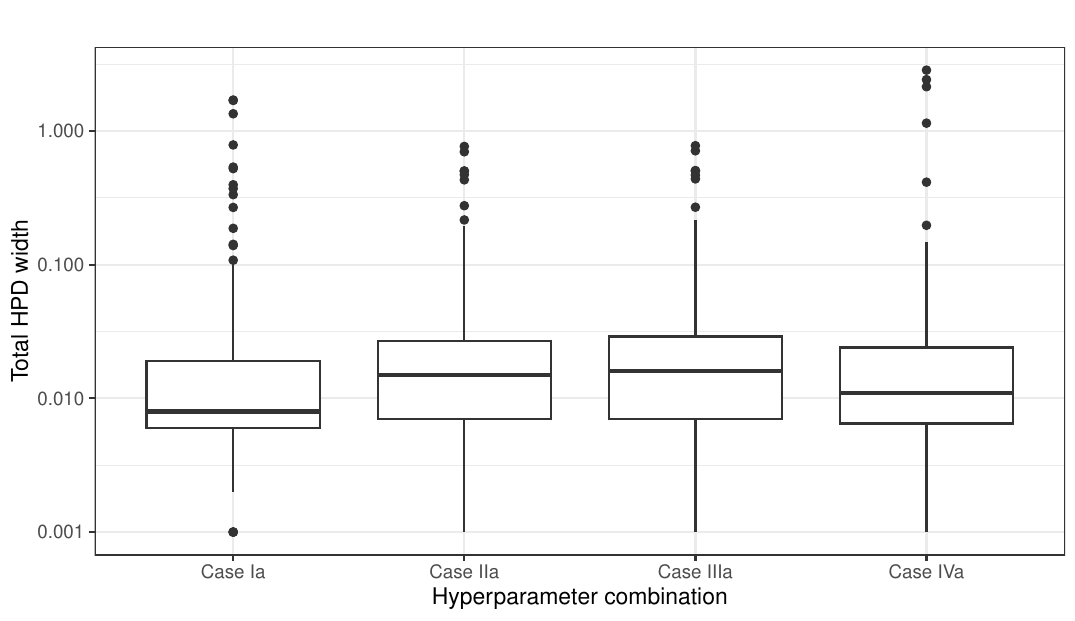}
    \caption{Each histogram depicts the total HPD width (on the log scale) for a different hyperparameter setting for the signal intensities.}
    \label{fig:hpd_widths_etas}
\end{figure}

\begin{figure}[ht]
    \centering
    \includegraphics[width=\linewidth]{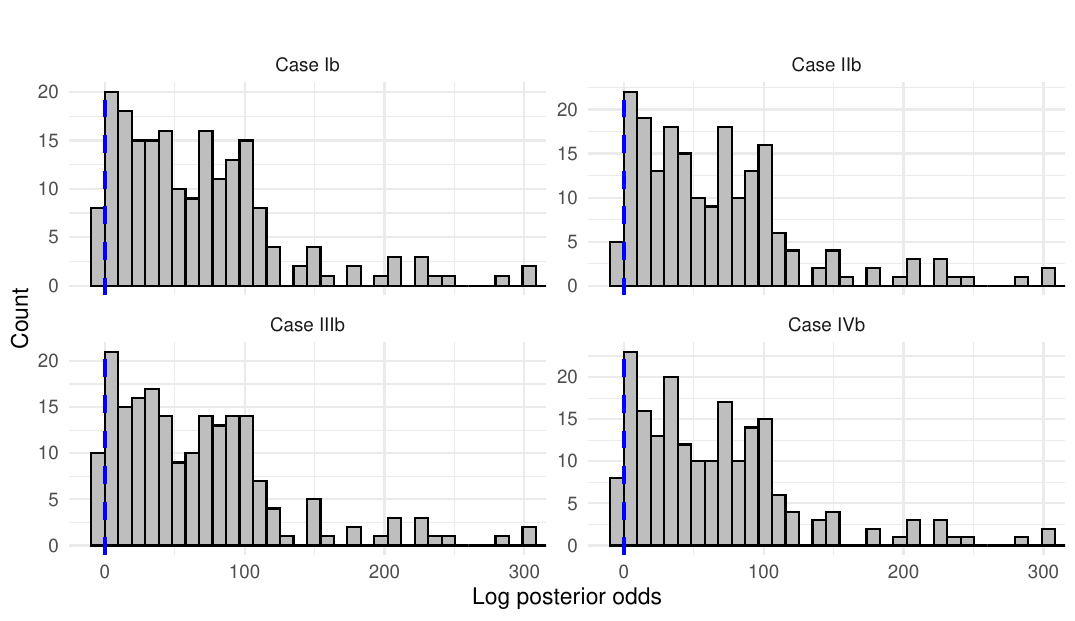}
    \caption{Each histogram depicts the log posterior odds for a different hyperparameter setting for redshift parameter. A small number of large log posterior odds values are cutoff from each plot. The dashed blue line at zero indicates the cutoff threshold.}
    \label{fig:log_PO_hists_zetas}
\end{figure}

\begin{table}[ht]
\centering
\begin{tabular}{lcc}
\hline
Setting & Rejection rate (Signal) & FPR (Null) \\
\hline
Ib & 0.962 & 0.019 \\
IIb & 0.977 & 0.033 \\
IIIb & 0.953 & 0.009 \\
IVb & 0.962 & 0.047 \\
\hline
\end{tabular}
\vspace{6pt}
\caption{Rejection probabilities on JWST data labeled as containing a signal by astronomers and the false positive rate (FPR) on simulated null data. Sample size of $n = 213$ per setting; maximum Monte Carlo standard error $\le 0.015$.}
\label{table:sens_zetas}
\end{table}

\subsubsection{Impact on redshift distributions}
Figure~\ref{fig:redshift_hists_etas} shows histograms of the marginal MAP redshift estimates under each prior specification, for all 213 spectra, including non-detections. The signal intensity prior has a pronounced effect on the overall redshift distribution. Cases Ia and IVa produce broadly similar patterns, though Case IVa yields fewer redshifts near zero and four. In contrast, Cases IIa and IIIa exhibit a substantial decrease in redshifts near two, with many more redshifts at lower and higher redshifts. This sensitivity is practically important. Relaxing prior information about relative emission line strengths alters which line configurations are favored, directly affecting the inferred redshift modes.

\subsubsection{Impact on redshift HPD set widths}
Figure~\ref{fig:hpd_widths_etas} shows boxplots for the total 99.865\% HPD set widths (on the log scale). Cases Ia and IVa again exhibit similar behavior, with Case IVa showing a modest increase in median width. Cases IIa and IIIa display larger median HPD set widths overall, indicating greater posterior uncertainty in redshift. Interestingly, extreme HPD widths are somewhat reduced in these cases relative to Cases Ia and IVa. Taken together, the HPD set results confirm that redshift uncertainty is sensitive to the signal intensity prior, though the magnitude of this sensitivity is less dramatic than its effect on detection decisions and the resulting marginal MAP estimates for redshift. 

\begin{figure}[ht]
    \centering
    \includegraphics[width=\linewidth]{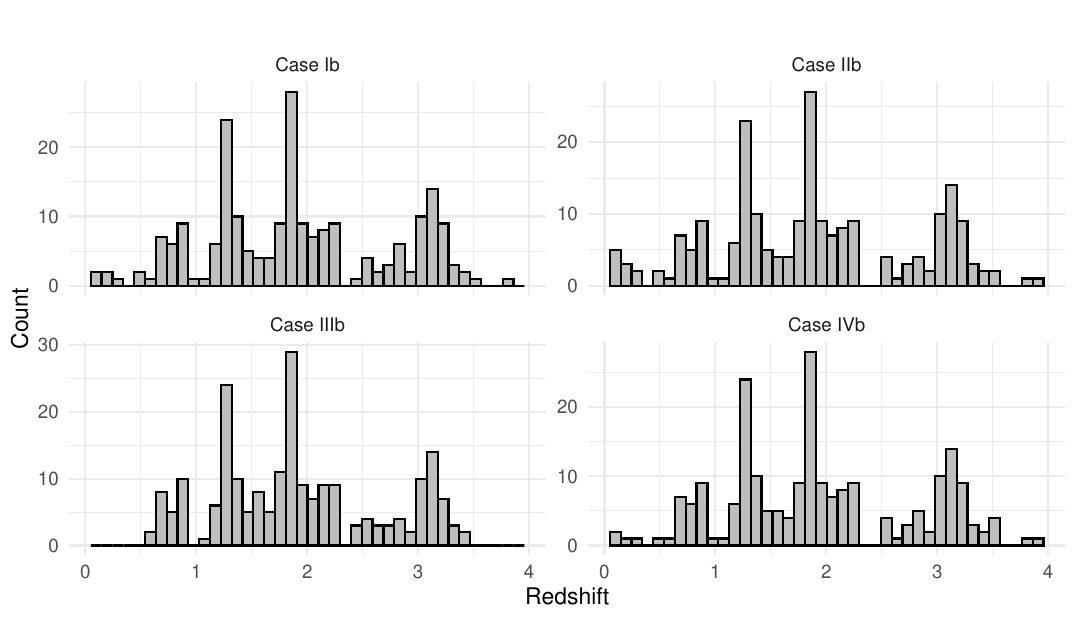}
    \caption{Each histogram depicts the distribution of marginal MAP redshift estimates for a different hyperparameter setting for redshift parameter.}
    \label{fig:redshift_hists_zetas}
\end{figure}

\begin{figure}[ht]
    \centering
    \includegraphics[width=\linewidth]{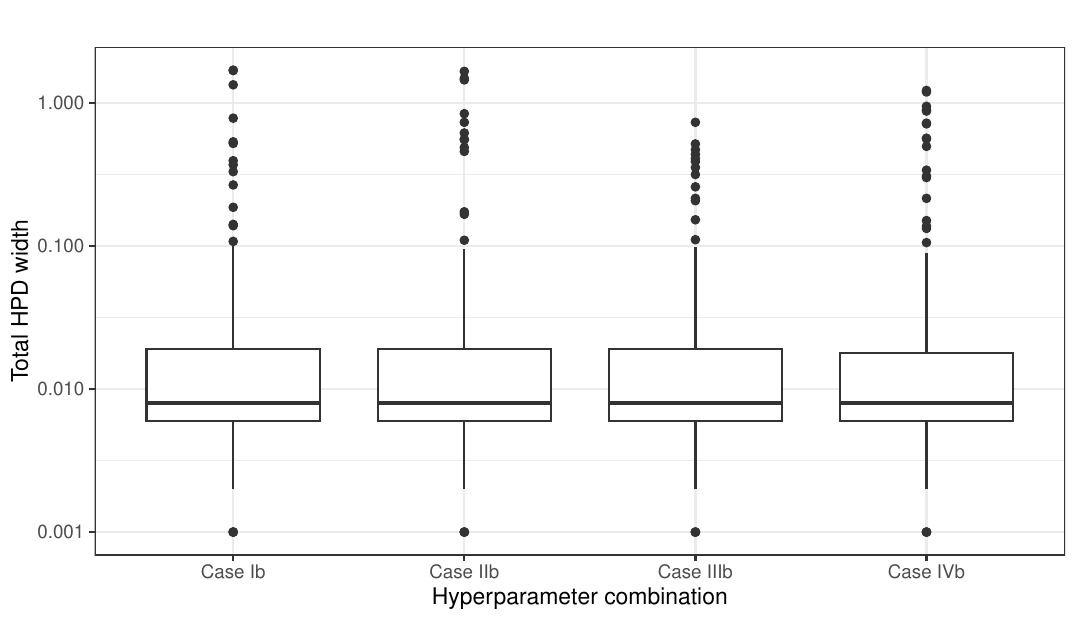}
    \caption{Each histogram depicts the total HPD width (on the log scale) for a different hyperparameter setting for redshift parameter.}
    \label{fig:hpd_widths_zetas}
\end{figure}

\subsection{Redshift hyperparameter sensitivity}
\label{app:redshift}
For redshift, we consider four scaled Beta priors over the interval $(0, 4]$:
\begin{itemize}
    \item Case Ib - Science-informed: $a_3 = 3, b_3 = 3$
    \item Case IIb - Uniform: $a_3 = 1, b_3 = 1$
    \item Case IIIb - Tightly concentrated symmetric: $a_3 = 10, b_3 = 10$
    \item Case IVb - Skewed: $a_3 = 3, b_3 = 1$.
\end{itemize}

\subsubsection{Impact on log posterior odds and FPRs}
Figure~\ref{fig:log_PO_hists_zetas} shows histograms of the log posterior odds under the four prior specifications. The overall shapes of the distributions are qualitatively similar, though the number of spectra identified as containing emission lines varies from 203 to 208 across settings. 

Table~\ref{table:sens_zetas} reports the rejection rates on JWST data and corresponding FPRs on simulated data. Both quantities exhibit some variation across prior choices, but the magnitude of sensitivity is small relative to that observed for the background or signal intensity hyperparameters. In particular, the largest FPR across the four cases remains below 0.047, indicating that emission line detection is relatively stable under these prior variations.

\subsubsection{Impact on redshift distributions}
Figure~\ref{fig:redshift_hists_zetas} shows histograms of the marginal MAP redshift estimates across the 213 spectra. The primary differences across prior settings occur in the tails of the redshift estimate distribution, particularly near zero and four. Case IIIb places essentially no prior mass near the boundaries, and accordingly yields no MAP estimates near zero or four. The remaining cases permit more extreme redshift values. Notably, the distribution under Case IIIb resembles the empirical distribution of astronomer-reported redshifts more closely than the other settings. Overall, aside from the boundaries, the distribution of the majority of galaxies remains broadly similar across prior specifications. 

\subsubsection{Impact on redshift HPD set widths}
Figure~\ref{fig:hpd_widths_zetas} displays boxplots of total 99.865\% HPD set widths (on the log scale) for redshift under each prior. Sensitivity of HPD widths to the redshift prior is minimal. The boxplots are nearly indistinguishable across cases, with differences appearing primarily in values near the boundaries. Cases Ib, IIb, and IVb each contain at least one spectrum with total HPD width exceeding one, while Case IIIb exhibits slightly smaller extremes, consistent with its reduced prior mass near the boundaries (that is, there are fewer plausible redshift values under this prior). Overall, the HPD set widths are largely stable to these changes in prior hyperparameters.